\documentclass[11pt,twoside]{book} 
\usepackage{asp2008n}
\usepackage{times}
\usepackage{lscape}
\usepackage{epsf}

\usepackage{graphicx}
\usepackage{amssymb}
\usepackage{longtable}
\usepackage[figuresright]{rotating}
\usepackage{multirow}

\defcitealias{ybd97}{YBD97}
\defcitealias{smz02}{SMZ02}

\newcommand{\simless}{\mathbin{\lower 3pt\hbox {$\rlap{\raise 5pt\hbox{$\char'074$}}\mathchar"7218$}}}

\mainmatter

\newlength{\deftabcolsep}
\begin{document}

\title{The Orion Molecular Cloud 2/3 and NGC~1977 Regions}   
\author{Dawn E. Peterson}   
\affil{Harvard-Smithsonian Center for Astrophysics\\
60 Garden Street, Cambridge, MA 02138, USA}

\author{S. Thomas Megeath}
\affil{Ritter Observatory MS 113, University of Toledo\\
2801 W. Bancroft St, Toledo, OH 43606, USA}

\begin{abstract}
The Orion Molecular Cloud 2/3 region (hereafter, OMC-2/3) and the
reflection nebula NGC~1977 encompass a section of the Orion A
molecular cloud undergoing vigorous star forming activity.  One of the
richest assemblages of protostars in the nearest 500 pc is seen in
OMC-2/3, while NGC 1977 contains a cluster of over 100 young stars.
In this review, we present a census of the protostars, pre-main
sequence stars, and young brown dwarfs in these regions.  These are
identified through sub-millimeter surveys, far-red to near-infrared
imaging and spectroscopy with ground-based telescopes, mid-infrared
photometry from the {\it Spitzer Space Telescope}, and X-ray
observations made with the Chandra X-ray Observatory.  We present an
overview of the distribution of molecular gas associated with these
regions and the rich complex of shock heated nebulae created by the
young stars interacting with the molecular gas.  Finally, we discuss
the relationship of OMC-2/3 and NGC~1977 to the neighboring Orion
Nebula Cluster and the Orion OB1 association.


\end{abstract}



\section{Introduction}

The section of the Orion A molecular cloud directly north of the Orion Nebula
is a region of intense star formation activity.  Although the number and
density of young stars in this area does not rival those found in the Orion
Nebula, it is still one of the most active regions of ongoing star formation
within 500 pc of the Sun.  It is typically divided into two regions: the Orion
Molecular Cloud 2/3 region, which contains a filamentary molecular cloud
rapidly forming new stars, and NGC~1977, a composite reflection and emission
line nebula.  In this chapter, we will review the observations
which delineate the stellar, sub-stellar, protostellar and gas content
in these two regions. We will begin by discussing the distribution of
stars, gas and dust in these two regions and their relationship to the
Orion Nebula Cluster.

\section{\hspace{-2ex}The Large Scale Distribution of Stars, Gas and Dust in OMC-2/3 and NGC~1977}

\begin{figure}[!ht]
\includegraphics[draft=False]{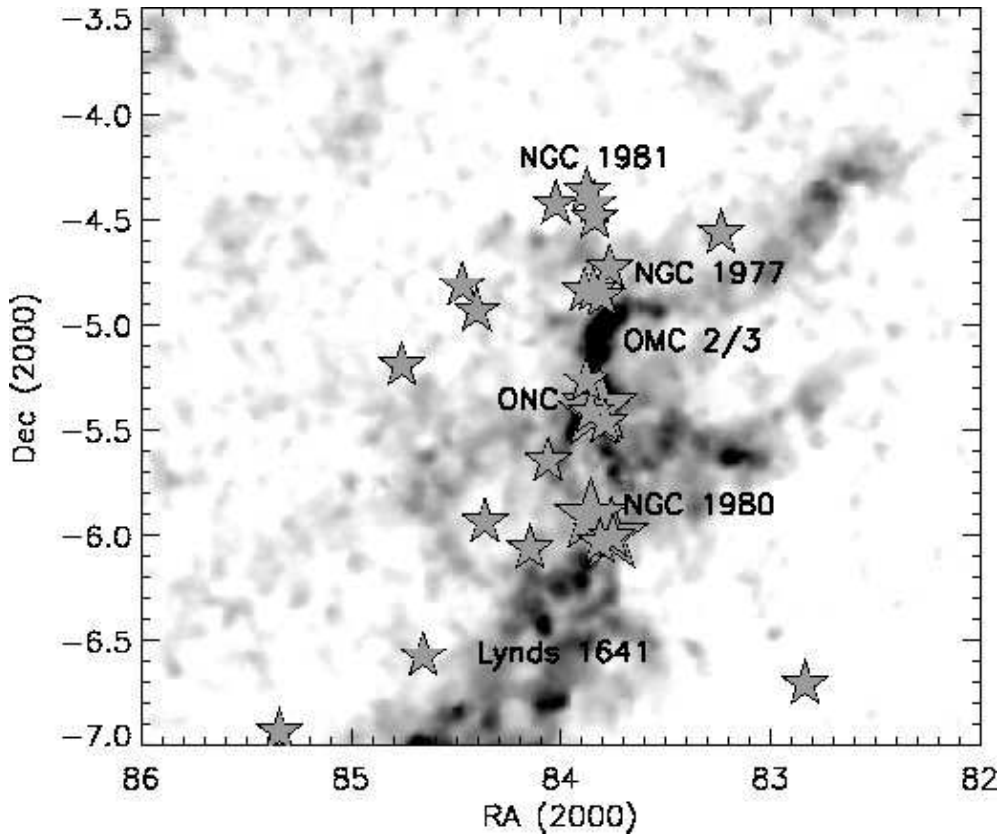}
\caption{The distribution of OB stars surrounding the OMC-2/3 and NGC~1977
regions.  The greyscale image is an extinction map of the Orion~A
cloud generated from the 2MASS PSC (Gutermuth, private communication).  The
star shaped markers are the OB stars in the OB 1 association \citep{b96phd};
the small markers are the B stars and the large markers are the O stars.
These show that the OMC-2/3 and NGC~1977 regions are surrounded by a number
of B stars, particular to the north of NGC~1977 and the east of OMC-2/3.
These OB stars are likely members of the 1c subgroup.}
\label{fig:obassoc}
\end{figure}

The Orion Molecular Cloud 2/3 (hereafter OMC-2/3), NGC~1977 and the Orion
Nebula region are  part of the Orion OB 1 complex (see Figure
\ref{fig:obassoc}); these regions fall along the line-of-sight of the Orion 1c
(4.6~Myr) and 1d ($<$ 1~Myr) subgroups \citep{b94}.  There are 15 OB stars in
the subgroup 1c and 18 OB stars in subgroup 1d \citep{orionbook}.
\citet{wh78} place the Trapezium stars of the Orion Nebula in the Orion 1d
association and the most massive star in NGC~1977 in Orion 1c.  The division
of stars into subgroups has been done primarily by their spatial distribution,
yet there may be significant spatial overlap between the subgroups (see, for
example, \citet{jmon06}).  Thus, the membership of the two subgroups and the
relationship of the subgroups to the current star formation in OMC-2/3 is far
from clear.  Surrounding OMC-2/3 and NGC~1977 are a number of B stars.
Directly north of NGC~1977, and outside the molecular cloud, is the NGC~1981
cluster which contains four B stars.  To the east of OMC-2/3 are three B
stars.  The relationship of these stars to the OMC-2/3 and NGC~1977 regions is
not understood; it is plausible that they have influenced star formation in
these regions by compressing, heating and/or ionizing the molecular gas in the
Orion~A cloud.

\begin{figure}[!tbp]
{\centering
\includegraphics[width=0.9\textwidth,draft=False]{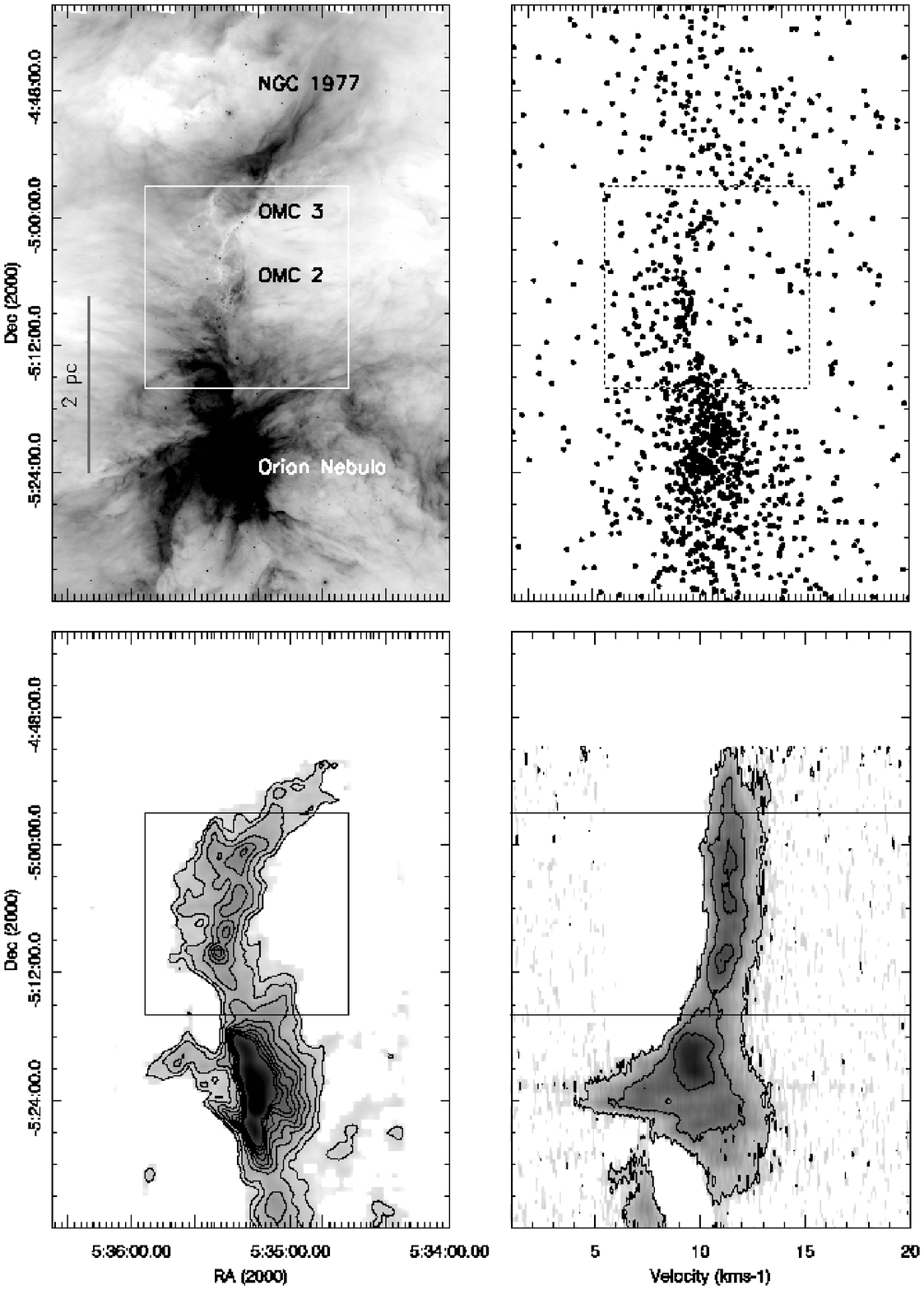}
\caption{OMC-2/3, NGC~1977, and the Orion Nebula Cluster.  The upper
  left panel shows 8~$\mu$m emission observed with IRAC on {\em
  Spitzer}.  The box shows the $20' \times 20'$ field mapped by
  \citet{peterson2007}.  The
  upper right panel shows the distribution of young stars and
  protostars identified by variability or infrared excess \citep*[][in
  prep]{chs01,megeath2007a}.  This map is incomplete toward the center
  of the Orion Nebula cluster.  The lower left panel shows the
  distribution of velocity integrated CS $2 \rightarrow 1$ emission
  mapped by Five College Radio Astronomy Observatory (N. Ridge,
  priv. comm.).  The lower right panel shows a radial
  velocity versus declination plot of the same data cube integrated
  over RA.  The two lines show the extent of the box shown in the
  other panels.}
\label{fig:tomfig_4panel}
}\end{figure}

In Figure \ref{fig:tomfig_4panel}, we show the relationship of OMC-2/3 and
NGC~1977 to the Orion Nebula using the positions of known young stars and
protostars and maps of the dust and gas emission.  The {\it Spitzer} 8~$\mu$m
map shows the emission of UV heated small grains and hydrocarbons.  The Orion
Nebula is the region of bright 8~$\mu$m emission in the southern part of the
map.  Extended diffuse emission is detected in OMC-2/3.  Within this emission,
dense molecular filaments are apparent in absorption against the diffuse
emission.  This filamentary cloud of dense molecular gas is seen in the lower
angular resolution CS~$(2 \rightarrow 1)$ map.  Apart from a shift in the
molecular gas velocity, as shown in the declination-velocity map, the gas in
the OMC-2/3 region appears to be a continuation of the same gas associated
with the Orion Nebula.  NGC~1977 is surrounded by a loop in the 8~$\mu$m
emission; the emission is brightest where the loop meets the dense molecular
gas.

Two studies have identified the pre-main sequence and
protostellar sources throughout the entire region displayed in Figure
\ref{fig:tomfig_4panel}.  \citet*{chs01} used multi-epoch $JHK_S$-band
imaging of a $0.84^\circ \times 6^\circ$ region centered on the Orion
Nebula to examine variability in young stars.  Variable stars in this
sample are likely to be young pre-main sequence stars.  More recently,
\citet[][in prep]{megeath2007a} have identified infrared excess sources
in the Orion A molecular cloud by combining mid-infrared photometry from the
{\it Spitzer} Space Telescope with the Two Micron All Sky Survey (2MASS)
point source catalog \citep{skrutskie06}; these sources are young
pre-main sequence stars and protostars with dusty disks and envelopes.

The observations show a continuous distribution of young stars
extending from the Orion Nebula, where the density of young stars is
highest, out to the NGC~1977 region.  The region of peak density is
often referred to as the Orion Nebula Cluster (hereafter, ONC, see the chapters by Muench et al. and O'Dell et al.).  From
the distribution of gas and stars, it is not clear where the ONC ends.
\citet{hill98} set an outer radius of the ONC of 2.06 pc, which
partially incorporates OMC-2.  \citet{carp2000} used the surface
density of 2MASS sources on the sky, corrected for background contamination,
to map the distribution of young stars in the Orion A cloud.  Using a
smoothed surface density plot, \citet{carp2000} identified clusters as
contiguous areas of elevated surface density; this analysis led to the
incorporation of the OMC-2/3 and NGC~1977 regions into a single cluster
centered on the Orion Nebula, four parsecs in radius, with over 1700 stars.
More recently, \citet[][in prep; see also \citet{allen2007}]{megeath2007a}
used the path linkage criterion of \citet{bat91} with a critical length of
0.33 pc to map the extent of the Orion Nebula cluster; they also include
OMC-2/3 and NGC~1977 in the ONC.

In summary, the OMC-2/3 and NGC~1977 regions appear to be part of a
complex of young stars and gas extending three parsecs from the Orion
Nebula.  These regions can be thought of as composing a continuous cluster
of young stars.  Observationally, these three regions are best
differentiated by the morphologies of their 8~$\mu$m nebulosity.  The
Orion Nebula is characterized by bright emission due to OB stars
heating the surface of the Orion A cloud; in this region, the
molecular cloud appears to be behind the bright nebulosity. In
contrast, the OMC-2/3 region shows the molecular gas in absorption against
extended diffuse emission.  Finally, the NGC~1977 region is surrounded
by a loop of emission; presumably the walls of a cavity cleared in the
molecular gas by the young B stars.  The distinctive morphologies
reflect the different environmental conditions in these three regions: the
ONC region containing young stars and molecular gas irradiated by intense
extreme UV radiation, the OMC-2/3 region containing protostars embedded
in molecular filaments, and the NGC~1977 region being largely cleared of
molecular gas and containing young stars irradiated by far-UV radiation from
three B stars.  These different environmental conditions are perhaps the best
motivation for dividing the northern end of the Orion A cloud into three
distinct regions.

\section{The Orion Molecular Cloud 2/3 Region}

\begin{figure}[t!]
\centering
\includegraphics[width=0.9\textwidth,draft=False]{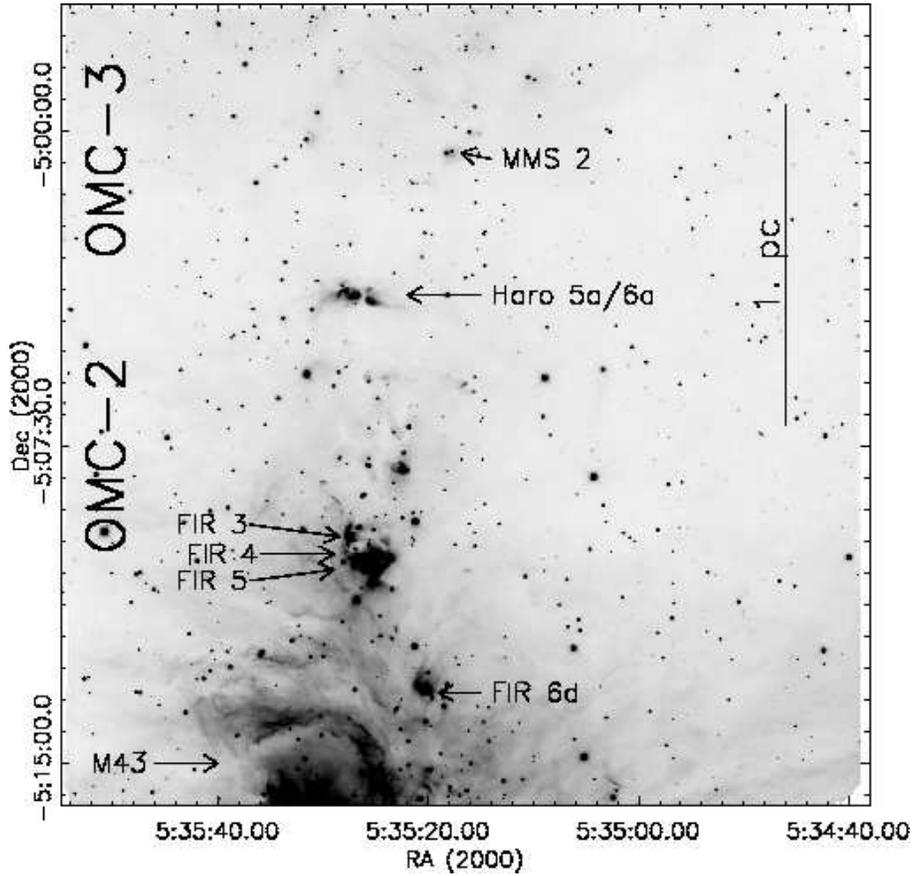}
\caption{OMC-2/3.  The $20' \times 20'$ $K-$band field mapped by
\citet{peterson2007} in their near-infrared study of OMC-2/3.  Sources
discussed in the text which are prominent in this image are labeled.}
\label{fig:omc23image}
\end{figure}

\begin{figure}[t]
\centering
\includegraphics[width=0.9\textwidth,draft=False]{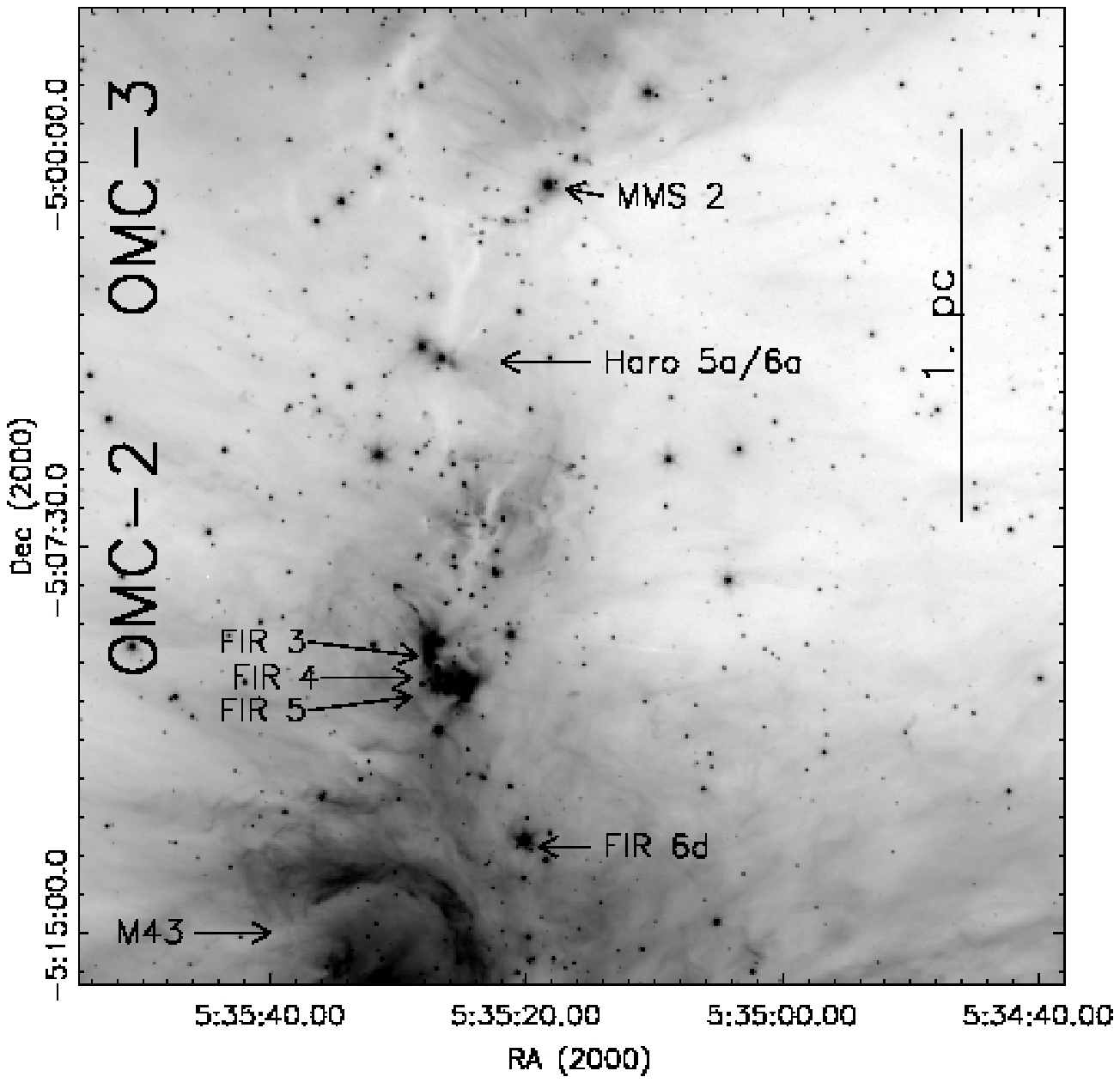}
\caption{OMC-2/3.  A $20' \times 20'$ image showing 4.5~$\mu$m band emission
observed with the IRAC camera on board the {\it Spitzer} Space Telescope.
This is the same region shown in Figure \ref{fig:omc23image}, and the
annotations from Figure \ref{fig:omc23image} are displayed.}
\label{fig:omc23image4.5}
\end{figure}

The OMC-2/3 region (see Figures \ref{fig:omc23image} and
\ref{fig:omc23image4.5}) encompasses a section of the Orion A molecular cloud
undergoing vigorous star forming activity, and at a distance of 450 pc
\citep{gs89}, it contains one of the richest assemblages of protostars in the
extended solar neighborhood.  It is important to note that four recent and
independent estimates of the distance to the ONC place it closer, consistent
with a value of 420 pc \citep{h07,jef07,mrfb07,sand07}; however, throughout
this discussion we use 450 pc.

OMC-2/3 is located in a molecular filament extending northward from the Orion
Nebula (NGC 1976) to NGC~1977.  Interest in OMC-2/3 was elevated by the
detection of a chain of 21 submillimeter condensations by \citet{chini97} and
the detection of numerous knots of H$_2$ emission from multiple outflows
\citep*{ybd97}.  These observations revealed for the first time the full
extent of the star formation activity in OMC-2/3.

The OMC-2 region was first defined by \citet{gbmnps74} in their
discovery of a group of near-infrared sources embedded in a molecular
core, centered at a position of $\alpha$ = 05$^h$ 35$^m$ 26.8$^s$,
$\delta$ = $-$05$^{\circ}$ 10$^{\prime}$ 17\arcsec (J2000).
The far-infrared luminosity of OMC-2 is estimated to be 2 $\times$
10$^{3} L_{\odot}$ \citep{thronson78}.  OMC-2 is north of OMC-1, which
is the molecular cloud associated with the Orion Nebula.
Interestingly, the discovery of OMC-3 is not well documented in the
literature.  In \citet{mztp74}, reference is made to a secondary
maximum 11$^{\prime}$ north of OMC-2 in a map of HCN $J = 1
\rightarrow 0$ emission that was possibly an artifact due to the
large, 2$^{\prime}$ beamwidth.  Two years later, \citet*{ket76} also
appear to detect a maximum north of OMC-2 in 2 millimeter H$_2$CO
emission at the same beamwidth, but at a distance of 16$^{\prime}$.
The first actual reference to ``OMC-3'' occurs in \citet{thronson78}
where far-infrared observations of OMC-3 were centered approximately
on $\alpha$ = 05$^h$ 35$^m$ 10.3$^s$, $\delta$ = $-$04$^{\circ}$
55$^{\prime}$ 01\arcsec (J2000), which is almost
16$^{\prime}$ north of OMC-2.  \citet{orionbook} also notes that OMC-3
is 16$^{\prime}$ north of OMC-2; a reference is not given but
presumably it is based on this observation.  Since that time, most of
the references to OMC-3 have referred to the location of the secondary
maximum in the HCN map from the initial \citet{mztp74} paper of
11$^{\prime}$ north of OMC-2.  In particular, \citet{chini97} define OMC-3 as
the region between 3$^{\prime}$ and 11$^{\prime}$ north of the center of OMC-2.

In the remainder of this section, we will define the OMC-2/3 as the gas
and associated stars within the $20' \times 20'$ field outlined in
Figure \ref{fig:tomfig_4panel}.  This is the region mapped by
\citet{peterson2007} in their deep, near-infrared observation of OMC-2/3
(Figures \ref{fig:omc23image} and \ref{fig:omc23image4.5}), and encompasses
the protostellar and prestellar sources detected in the submillimeter surveys.

\subsection{Molecular Gas}

The \citet{b87} Bell Labs survey of $^{13}$CO $1 \rightarrow 0$ emission
extensively mapped the Orion A molecular cloud covering 8 deg$^{2}$; the total
mass was found to be 5$\times$10$^4 M_{\odot}$.  The map shows an
integral shaped filament in the northern part of the Orion A cloud from NGC
1977 in the north through the Orion Nebula in the middle down to about
20$^{\prime}$ north of NGC 1999 in the south.  This filament, which includes
OMC-2/3, is about 0.5 $\times$ 13 pc long and has a total mass of
5$\times$10$^3 M_{\odot}$, or about 25\% of the total mass of their map of
the Orion A cloud.  \citet{b87} suggest that the filament results from the
interstellar medium being compressed by a superbubble driven by the Orion OB
association.

\citet{dutrey93} obtained J = $2 \rightarrow 1$ C$^{18}$O maps of the Orion A
cloud and found that the filament seen in the \citet{b87} map breaks up into
several discrete fragments.  From these observations, the molecular gas masses
of OMC-2 and OMC-3 are found to be 113$M_{\odot}$ and 140$M_{\odot}$,
respectively.  However, a higher resolution study by \citet{cl95} in several
transitions of $^{13}$CO, C$^{18}$O, C$^{32}$S and C$^{34}$S shows both these
regions to be much more complex.  \citet{cl95} shows a single dense core in
OMC-2 with a mass of 22$M_{\odot}$ and a radius of 0.11 pc.  In contrast,
they find the gas traced by C$^{18}$O and CS in OMC-3 is highly elongated with
a total mass of 55$M_{\odot}$.  The CS observations show three dense cores in
OMC-3 with masses of 12, 10 and 13$M_{\odot}$; since CS traces only the dense
gas, it is inferred that the C$^{18}$O maps are detecting a massive envelope
surrounding the dense cores.

\subsection{Submillimeter Sources}

Submillimeter observations of OMC-2/3 have been conducted at 1300
$\mu$m \citep*{mwz90,chini97, ncm03}, 850 $\mu$m, 450 $\mu$m
\citep{jb99}, and 350 $\mu$m \citep{lis98}, resulting in the discovery
of a wealth of pre-stellar, Class 0 and Class I objects.  Twenty-one
compact sources were detected in OMC-2/3 by \citet{chini97} in their
1300 $\mu$m dust imaging of the region.  Previous to \citet{chini97},
the OMC-2 core was observed at 1300 $\mu$m by \citet*{mwz90} and their
original six sources, listed as FIR 1a, 2$-$5 and 6a in Table
\ref{table:submmsources}, were also detected by \citet{chini97}.  In
OMC-2, \citet{chini97} detected two additional sources north of FIR 1a
(FIR 1b and 1c) as well as a cluster of compact sources near FIR 6a:
FIR 6b$-$6e.  Finally, \citet{chini97} found four more sources in OMC-2 (MMS
7-10) and six sources in OMC-3 (MMS 1-6).  Based on L$_{bol}$/L$_{submm}$
ratios $<$ 200, all six condensations in OMC-3 fit the definition of Class 0
objects, and are thus in the earliest stages of protostellar evolution
\citep{chini97}.

\setlength{\tabcolsep}{0.8\deftabcolsep}
\begin{table}[!tpb]
\caption{Submillimeter Sources in OMC-2/3}
\smallskip
{
\scriptsize
\begin{tabular}{l@{\hskip3pt}c@{\hskip3pt}c@{\hskip3pt}c@{\hskip3pt}c@{\hskip3pt}c@{\hskip3pt}l}
\tableline
{\smallskip}
Source & R.A.\tablenotemark{a} & Dec.\tablenotemark{a} & S (350 $\mu$m) & S (1300 $\mu$m) & S (3.6 cm) & Other \\
Name\tablenotemark{b} & (J2000) & (J2000) & (Jy) & (mJy) & (mJy) & Designations\tablenotemark{b} \\
\noalign{\smallskip}
\tableline
\noalign{\smallskip}
MIR 1/2   & 05 35 28.3 &  $-$04 58 38 & & 290 & & \\
CSO 1  & 05 35 29.8 &  $-$04 58 47 & 15 & & &  \\
CSO 2  & 05 35 13.9 &  $-$04 59 25 & 15 & & &  \\
MIR 3  & 05 35 15.8 &  $-$04 59 59 & 17 & 179 & & CSO 3 \\
CSO 4  & 05 35 17.1 &  $-$05 00 03 & 19 & & & \\
MMS 1  & 05 35 18.1 &  $-$05 00 20 & 26 & 347 & & CSO 5 \\
MMS 2  & 05 35 18.5 &  $-$05 00 30 & 25 & 249 & 0.25 & CSO 6; VLA 1; \\
& & & & & & MIR 5/6 \\
MMS 3\tablenotemark{c}   & 05 35 19.2 &  $-$05 00 51 & 26 & 152 & & CSO 7 \\
MMS 4  & 05 35 20.5 &  $-$05 00 50.0 & 27 & 354 & & CSO 8; MIR 7  \\
MMS 5  & 05 35 22.5 &  $-$05 01 15 & 24 & 400 & & CSO 9 \\
MMS 6 & 05 35 23.5 &  $-$05 01 31 & 45 & 2000 & 0.15 & CSO 10; VLA 3 \\
CSO 11 & 05 35 27.1 &  $-$05 03 39 & 15 & 164 & & MIR 9 \\
MMS 7 & 05 35 26.5 &  $-$05 03 57 & 19 & 360 & 0.59 & CSO 12; VLA 4; MIR 10; \\
& & & & & & IRAS 05329-0505 \\
MMS 8 & 05 35 26.7 &  $-$05 05 19 & 17 & 268 & & CSO 13  \\
MMS 10 & 05 35 32.3 & $-$05 05 42 & & 166 & & \\
MMS 9 & 05 35 26.2 &  $-$05 05 44 & 21 & 412 & 0.79 &  CSO 14, VLA 5 \\
MIR 11 & 05 35 31.5 &  $-$05 05 47 & & 228 & & \\
MIR 12 & 05 35 25.8 &  $-$05 05 58 & & 146 & & \\
CSO 15 & 05 35 22.9 &  $-$05 06 41 & 14 & & & \\
FIR 1c & 05 35 23.8 &  $-$05 07 09 & 18 & 180 & & CSO 16; MIR 14; \\
& & & & & & IRAS 05329-0508 \\
FIR 1b & 05 35 23.7 &  $-$05 07 35 & 18 & 195 & & CSO 17 \\
FIR 1a & 05 35 24.4 &  $-$05 07 53 & 18 & 393 & 0.31 & CSO 18; VLA 8\tablenotemark{d} \\
CSO 19 & 05 35 25.0 &  $-$05 08 19 & 15 & & & \\
CSO 21 & 05 35 26.4 &  $-$05 08 26 & 14 & & & \\
FIR 2 & 05 35 24.4 &  $-$05 08 34 & 17 & 340 & & CSO 20; MIR 19 \\
MIR 20 & 05 35 26.9 &  $-$05 09 24 & & 351 & & \\
FIR 3 & 05 35 27.5 &  $-$05 09 38 & 36 & 676 & 2.48 & CSO 22; VLA 11; \\
& & & & & & MIR 21/22 \\
FIR 4 & 05 35 26.8 &  $-$05 10 03 & 67 & 1252 & 0.64 & CSO 23; VLA 12; MIR 24?;\\
& & & & & & IRAS 05329-0512 \\
MIR 25 & 05 35 27.6 &  $-$05 10 09 & & 216 & & \\
MIR 26 & 05 35 28.2 &  $-$05 10 14 & & 423 & & \\
FIR 5 & 05 35 26.4 &  $-$05 10 24 & 34 & 452 & & CSO 24; MIR 27? \\
MIR 28 & 05 35 24.8 &  $-$05 10 30 & & 186 & 1.04 & VLA 13\\
FIR 6b & 05 35 23.3 &  $-$05 12 09 & 15 & 300 & & CSO 25; MIR 29 \\
CSO 26 & 05 35 26.3 &  $-$05 12 23 & 14 & & & \\
CSO 27 & 05 35 24.7 &  $-$05 12 33 & 19 & & & \\
FIR 6a & 05 35 23.4 &  $-$05 12 42 & 21 & 361 & & CSO 28 \\
MIR 30 & 05 35 18.2 &  $-$05 13 07 & & 186 & & \\
FIR 6c & 05 35 21.6 &  $-$05 13 14 & 18 & 450 & & CSO 29 \\
FIR 6d & 05 35 19.8 &  $-$05 13 19 & 12 & 314 & & CSO 30; MIR 31/32 \\
MIR 36 & 05 35 22.6 &  $-$05 14 11 & & 184 & & \\
CSO 31 & 05 35 21.5 &  $-$05 14 30 & 15 & 189 & & MIR 37 \\
CSO 32 & 05 35 21.3 &  $-$05 14 54 & 15 & 161 & & MIR 38 \\
MIR 39/40 & 05 35 19.9 &  $-$05 15 09 & & 140 & & \\
MIR 41 & 05 35 23.5 &  $-$05 15 23 & & 42 & & \\
CSO 33 & 05 35 19.5 &  $-$05 15 35 & 20 & 288 & & MIR 42\\
MIR 43/44 & 05 35 25.3 &  $-$05 15 36 & & 48 & & \\
\noalign{\smallskip}
\tableline
\noalign{\smallskip}

\multicolumn{7}{l}{\parbox{0.9\textwidth}{\footnotesize $^a$~Positions from \citet{lis98} except for source 15 which has \citet{chini97} coordinates. For non-CSO or MMS sources, MIR coordinates are used \citep*{ncm03}.}}\\
\multicolumn{7}{l}{\parbox{0.9\textwidth}{\footnotesize $^b$~CSO names
and corresponding fluxes from \citet{lis98}, MMS and FIR from
\citet{chini97}, VLA from \citet*{rrc99} and MIR from
\citet*{ncm03}.  IRAS sources from the Point Source Catalog.  The
350 $\mu$m and 1300 $\mu$m positions agree to within $\lesssim$ 6\arcsec \citep{lis98}.}}\\[1ex]
\multicolumn{7}{l}{\parbox{0.9\textwidth}{\footnotesize $^c$~There is a positional difference between CSO 7 and MMS 3 of $\sim$ 19\arcsec; consequently, the association of these sources is uncertain.  Coordinates are from \citet{lis98}.}}\\[1ex]
\multicolumn{7}{l}{\parbox{0.9\textwidth}{\footnotesize $^d$~There is a positional difference between FIR 1a and VLA 8 of $\sim$ 8\arcsec; however, FIR 1a has a broad peak and VLA 8 is located within the boundary so the association is likely \citep*{rrc99}.}}\\[1ex]
\label{table:submmsources}
\end{tabular}
}
\end{table}
\setlength{\tabcolsep}{\deftabcolsep}

\begin{figure}[!tb]
\centering
\includegraphics[width=\textwidth]{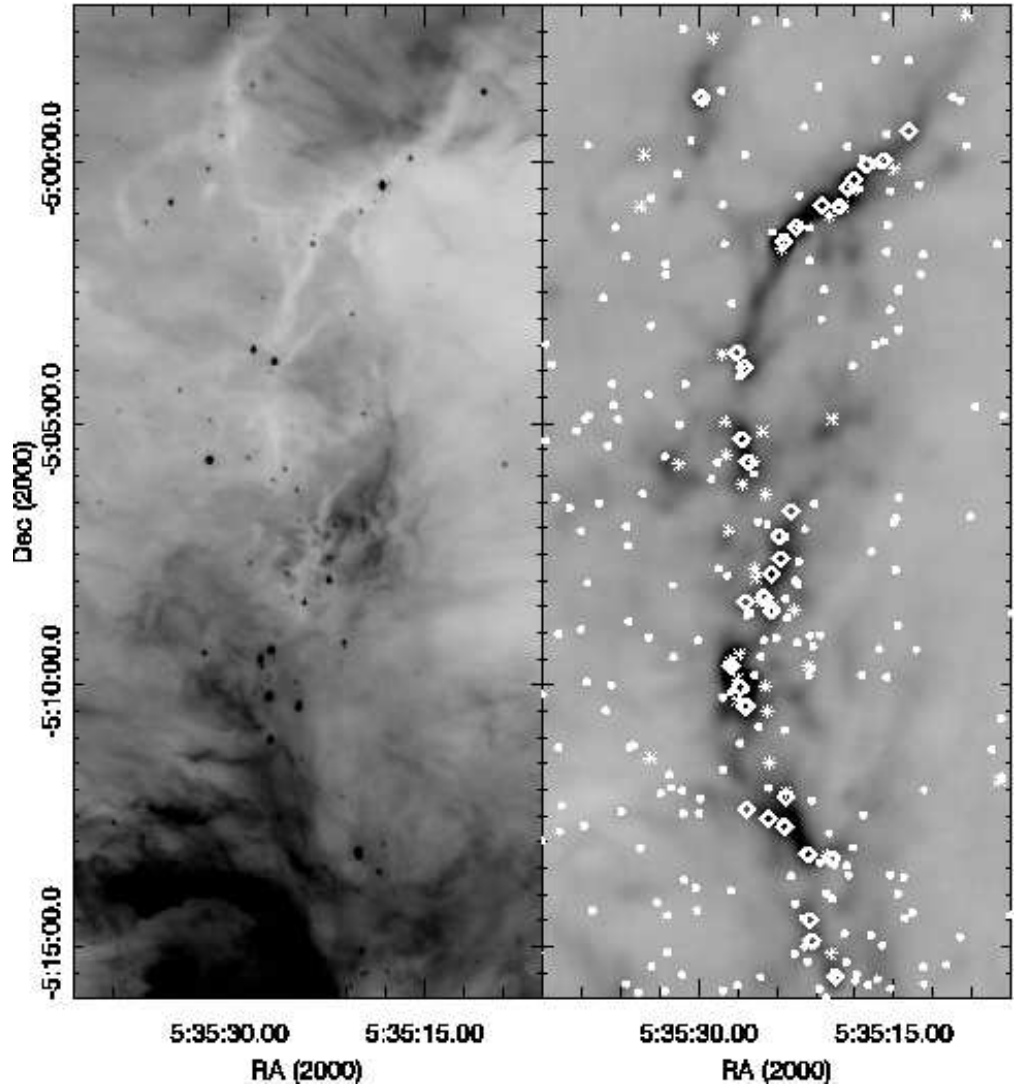}
\caption[Dense Gas + Protostars in OMC-2/3]{The central part of the
OMC-2/3 region shown in Figure \ref{fig:omc23image}.  The left panel shows
the {\it Spitzer} 8~$\mu$m map.  The right panel shows the 850~$\mu$m
map of \citet{jb99} in inverse greyscale. Overlaid on this map are the
following: 1.) positions of the 34 submillimeter sources given in Table
\ref{table:submmsources} (diamonds), 2.) positions of {\it Spitzer}
identified protostellar candidates (asterisks) and 3.) positions of pre-main
sequence stars (circles; see Section \ref{sec:spitzer}).  Note that many of
the filaments seen in emission in the 850~$\mu$m map are seen in absorption
against the diffuse background in the 8~$\mu$m map.}
\label{fig:tom_proto}
\end{figure}

All 21 sources of \citet{chini97} were detected at 350 $\mu$m by
\citet{lis98} except for one (MMS 10), and an additional 13 sources
were discovered (CSO 1$-$33 in Table \ref{table:submmsources}).  In
Figure \ref{fig:tom_proto}, these 34 submillimeter sources are
shown with the 8~$\mu$m IRAC map as well as the 850 $\mu$m map from
\citet{jb99}.  Overall, the submillimeter sources follow the
filamentary structure of the 850 $\mu$m emission.

\citet*{ncm03} also obtained 1300 $\mu$m maps of OMC-2/3 and identified 12 new
1300 $\mu$m peaks in addition to the ones already identified; all are listed
in Table \ref{table:submmsources} (MIR sources).  They complemented the
submillimeter data with mid-infrared imaging and detected at least 13 (and as
many as 15) of the submillimeter peaks at 11.9 $\mu$m ($N-$band) and 17.8
$\mu$m ($Q-$band).  In the mid-infrared, 6 of the submillimeter peaks are
resolved into double sources, e.g. MMS 2 from \citet{chini97} is resolved
into MIR 5 and 6 \citep*{ncm03}.  The projected spatial separation of these
two particular sources is 1.3\arcsec, or 570 AU at 450 pc, with a
position angle of 45$^{\circ}$.

Two of the submillimeter sources of \citet{chini97} coincide with Infrared
Astronomical Satellite (IRAS) sources: IRAS 05329$-$0505 (MMS 7) and IRAS
05329$-$ 0512 (FIR 4).  There are also two submillimeter sources detected in
X-rays: MMS 2 and 3 \citep{tsuboi01}.

Seven of the \citet{chini97} sources were also detected at 3.6 cm
\citep*{rrc99}.  They present radio continuum maps and detect a total of
14 sources at 3.6 cm, where 3 are likely background sources.  The remaining 11
are likely either young stars or protostars.  Of those 11 sources, 7
are also detected at 1300 $\mu$m and are listed in Table
\ref{table:submmsources} (VLA 1, 3, 4, 5, 8, 11, 12).  These centimeter
sources are thought to result from ionized gas in outflows, and they are
clear evidence that many of the submillimeter sources are indeed protostars
containing hydrostatic cores \citep*{rrc99}.

Submillimeter wavelength polarimetry of OMC-3 has shown that the magnetic
field is perpendicular to the filament along most of the length of the region
encompassed by the MMS 1-9 cores \citep{mw00}.  They also find that the
outflows in this region are not consistently aligned with the ambient field or
the filament.

\begin{figure}[!ht]
\plotfiddle{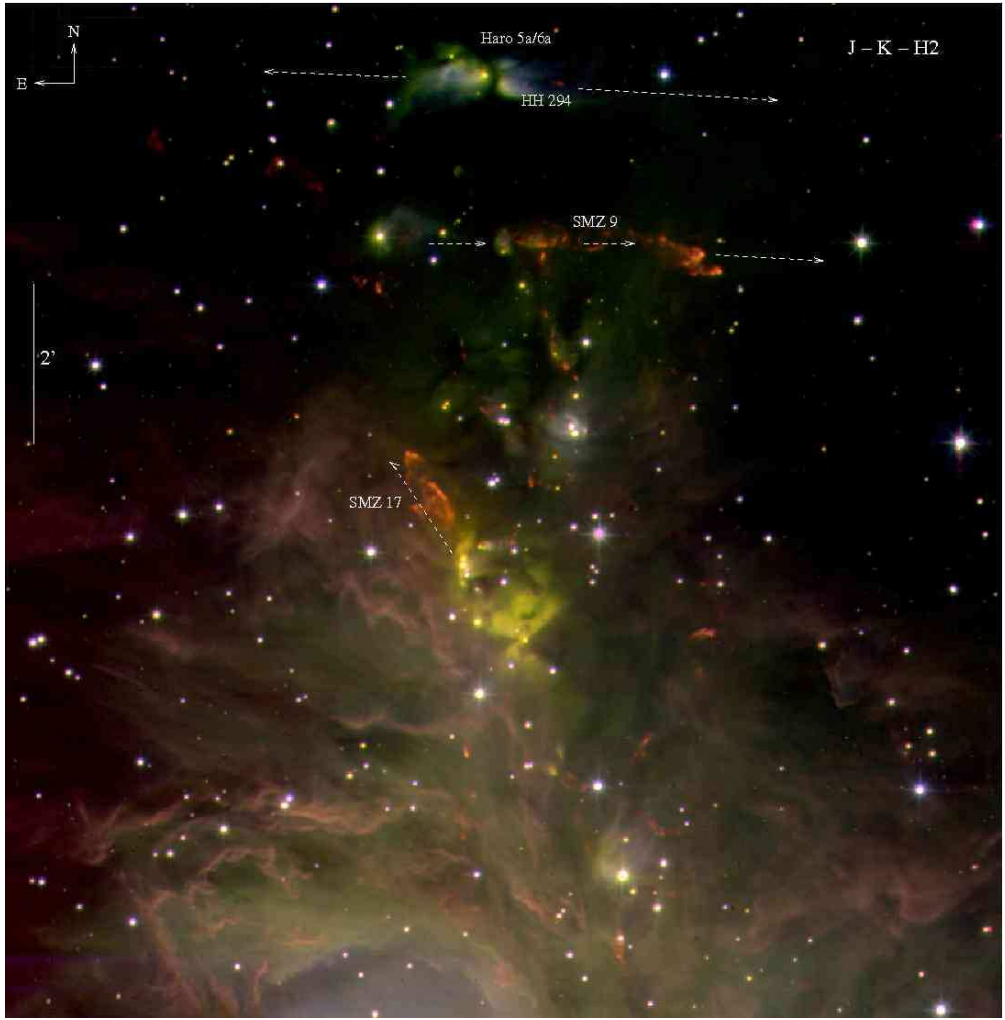}{13cm}{0}{48}{48}{-190}{0}
\caption[Outflows in OMC 2/3]{Near-infrared J-K-H$_2$ color-composite image of the OMC-2/3 region. Jets and outflows appear in red.  The flow associated with Haro 5a/6a is SMZ~6.  Data obtained with the wide-field imager, WFCAM, on UKIRT.  Figure courtesy of Chris Davis (private communication).}
\label{fig:outflows_chris}
\end{figure}

\subsection{Outflows \label{sec:outflows}}

\begin{figure}[!ht]
\plotone{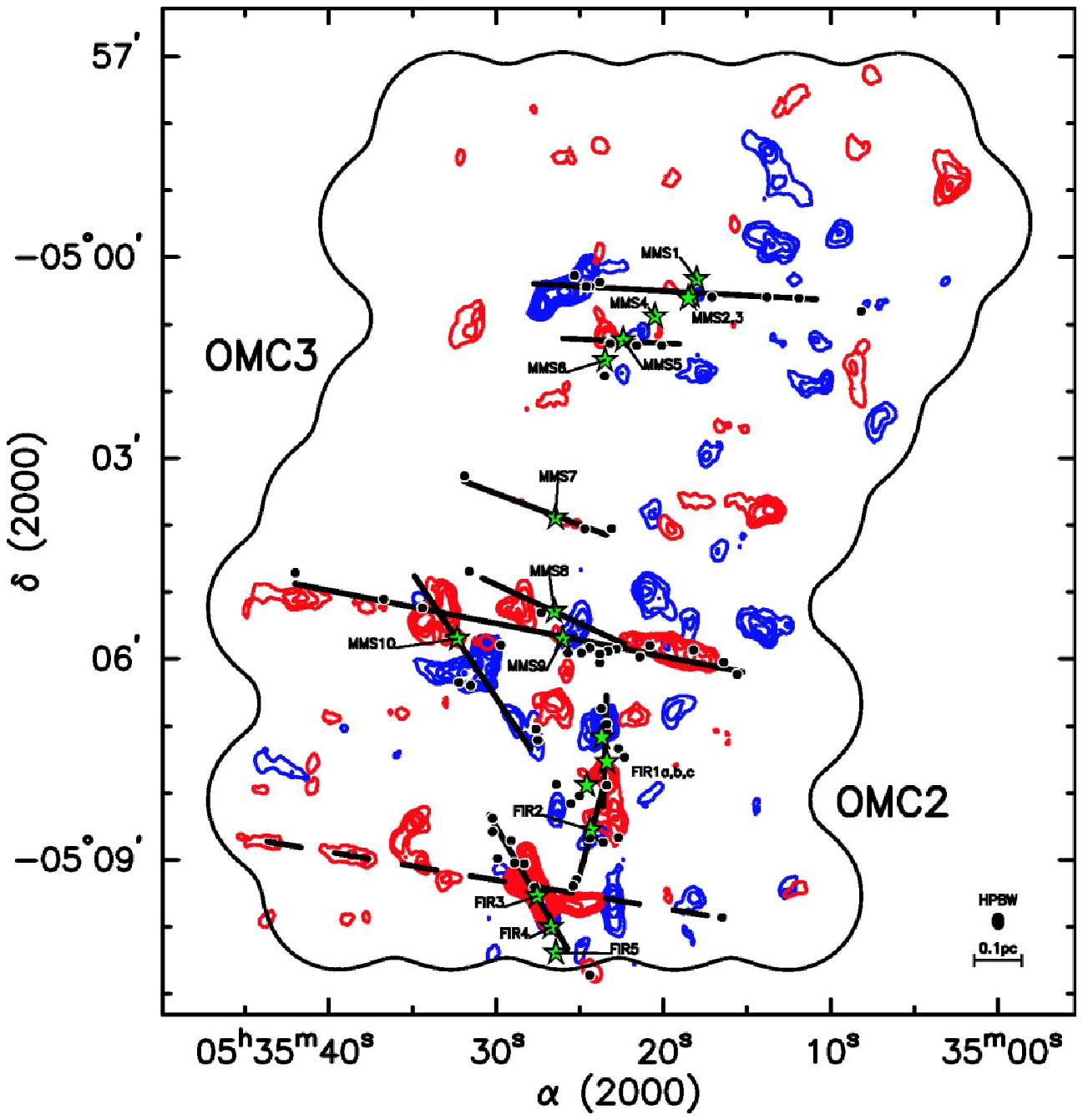}
\caption[Figure 2 from \citet*{wph03}]{This figure from \citet*{wph03} shows
line-wing emission from their BIMA mosaic (interferometer data only).  The
blue and red contours show emission integrated from $v =$ 5.5 to 9.0 km
s$^{-1}$ and 12.5 to 16.0 km s$^{-1}$, respectively.  The contour starting
levels and increments are the same (8, 4 K km s$^{-1}$) for both velocity
ranges.  The star symbols represent the positions of the \citet{chini97}
submillimeter sources and dots represent the H$_2$ knots from \citet*{ybd97}.
The solid contour around the map is the 30\% sensitivity level and the
8.8$^{\prime\prime} \times$ 13.4\arcsec\ beam is shown in the lower
right-hand corner.}
\label{fig:wph03}
\end{figure}

Infrared, optical and millimeter measurements have shown that OMC-2/3 is
rich with outflows driven by the young stars in this region.
\citet*[][hereafter, YBD97]{ybd97} discovered 80 knots from 2.12 $\mu$m
imaging of the $v = 1 - 0$ S(1) line of shock excited H$_2$ with nearly a
dozen of those being collimated outflows from young stellar objects (YSOs) in
OMC-2/3.  Observations in $^{12}$CO $J = 2\rightarrow1$ by \citet{yu00}
suggest that these H$_2$ knots are driving molecular outflows.  OMC-2/3 was
also observed at 2.12~$\mu$m by \citet*[][hereafter SMZ02]{smz02} and they
identified 24 flows labeled SMZ 2$-$25.  Tables 2 and 3 in \citetalias{smz02}
enumerate the flows (Table 3) and each of the H$_2$ knots associated with
those flows (Table 2); Table 2 also includes a comprehensive correlation of
the HH objects and \citetalias{ybd97} H$_2$ knots with the \citetalias{smz02}
H$_2$ features.  Using these positions, \citet*{wph03} identified nine
outflows in CO emission from high resolution Berkeley-Illinois-Maryland Array
(BIMA) imaging.  Figure \ref{fig:wph03}, from \citet*{wph03}, shows the
location of those nine outflows, each of which appears to originate from a
\citet{chini97} submillimeter source.  The nine outflows line up with
approximately 50 of the 60 \citetalias{ybd97} H$_2$ knots that are coincident
with the BIMA map; those that do not are likely to be outflows in the plane of
the sky \citep*{wph03}.  \citet{aso00} identified 4 additional molecular
outflows from CO ($1-0$) and HCO$^+$ ($1-0$) observations with the Nobeyama
45-m telescope, and all are associated with the known far-infrared or
millimeter sources \citep{chini97,lis98}.

The most prominent Herbig Haro objects in OMC-2/3 are associated with the
bipolar reflection nebula known as Haro~5a/6a, first noted by \citet{haro53}.
Haro~5a/6a appears almost edge-on (see Figures \ref{fig:omc23image},
\ref{fig:omc23image4.5}, \& \ref{fig:outflows_chris}), exhibiting a dark lane
between the two lobes of the nebula at optical and near-infrared wavelengths
\citepalias{ybd97,smz02}.  A slightly tilted obscuring torus is inferred from
imaging polarimetry data \citep{wol86}, while near-infrared spectroscopy
suggests that the source hidden by the torus may be an FU Orionis type
variable \citep{ra97,gar08}.

Haro~5a/6a is centered on IRAS source 05329-0505, which is one of the 1300
$\mu$m sources (MMS 7) detected by \citet{chini97}.  It is also associated
with a radio jet \citep{rrab04} and a large-scale bipolar outflow
\citep*{ybd97,wph03}.  The western lobe of the nebula (Haro~5a) includes a
knot of H$_2$ emission coinciding with HH~294 (see our Table
\ref{table:hhobjects} for a list of the prominent outflows in OMC-2/3), a
well collimated optical jet 100\arcsec\ in length which flows from
the IRAS source \citep*{rbd97}.  The flow axis intercepts more distant HH
knots
42,
128, and
295 \citep*{ogu91,rbd97}, although the complex web of
flows in this region makes concrete associations difficult -- particularly
at infrared and millimeter wavelengths \citep*[][in prep; see also
Figures~\ref{fig:outflows_chris} \&
\ref{fig:wph03}]{sch97,ybd97,yu00,smz02,wph03,dav07}.  Large, though faint,
H$_2$ bow shocks are detected by \citetalias{smz02}, which they associate with
flow SMZ~6 originating from Haro~5a/6a (also called flow ``F'' in
\citetalias{ybd97}).  This source has also been detected in 3$-$24 $\mu$m
{\it Spitzer} observations and is classified as a protostar on the basis of
the {\it Spitzer} photometry (see Section \ref{sec:spitzer}).

About 3\arcmin\ north of Haro~5a/6a, a compact chain of optical/infrared knots
trace HH~293 (flow SMZ~5 in \citetalias{smz02}; flow ``C'' in
\citetalias{ybd97}).   Bright H$_2$ bows delineate a third flow
$\sim$2\arcmin\ to the south of Haro~5a/6a (flow SMZ~9 in \citetalias{smz02};
flow ``H'' in \citetalias{ybd97}; labeled in Figure
\ref{fig:outflows_chris}).  A fourth H$_2$ flow (flow SMZ~3 in
\citetalias{smz02}; flow ``B'' in \citetalias{ybd97}) is detected a further
arcminute north-west of HH~293.  Notably, all four flows are more-or-less
parallel, and orthogonal to the chain of cores that runs roughly north-south
through the center of the region.  Further south toward FIR~4 and the region
directly west of M~43, a complex of star forming cores produces at least
another half-dozen H$_2$ flows \citep*{yu00,smz02,dav07}, although these are
more randomly oriented.  \citet{chini97}, \citet{yu00} and \citet*{wph03}
present molecular line maps of the outflows in this region.  The extremities
of many of the flows in this region, where they break out of the north-south
molecular filament that pervades OMC-2/3, are manifest as optically-visible HH
objects \citep*{ogu91,rbd97}.

\setlength{\tabcolsep}{0.95\deftabcolsep}
\begin{table}[!ht]
\caption{Prominent Herbig-Haro and H$_2$ outflows in OMC-2/3}
\smallskip
\begin{center} {\footnotesize
\begin{tabular}{l@{\hskip5pt}l@{\hskip5pt}c@{\hskip5pt}c@{\hskip5pt}c@{\hskip5pt}c@{\hskip5pt}c@{\hskip5pt}l}
\tableline
\noalign{\smallskip}
HH or H$_2$    & Driving  & RA\tablenotemark{c} & Dec\tablenotemark{c} & Other flow & Length & Refs\tablenotemark{e}\\
designation\tablenotemark{a}& source\tablenotemark{b}   & (2000) &  (2000)     & designations   &  (pc)\tablenotemark{d}  &  \\
\noalign{\smallskip}
\tableline
\noalign{\smallskip}

HH~294            & IRAS05329-0505;   & 05 35 26.4 & $-$05 03 54 & Haro~5a/6a;       & 1.43 & 1,2,3,4\\
                  & MMS~7             &            &             & SMZ~6; YBD ``F''  &      &\\
HH~293            & MMS~5             & 05 35 22.4 & $-$05 01 16 & SMZ~5; YBD ``C''  & 0.05 & 1,3,4\\
SMZ~9             & MMS~9             & 05 35 26.2 & $-$05 05 46 &  YBD ``H''        & 0.34 & 3,4\\
SMZ~3             & IRS~1             & 05 35 18.2 & $-$05 00 33 &  YBD ``B''        & 0.70 & 3,4\\
\noalign{\smallskip}
\tableline
\noalign{\smallskip}
\multicolumn{7}{l}{\parbox{0.9\textwidth}{\footnotesize $^a$~SMZ refers to the H$_2$ flow nomenclature of \citet*{smz02}.}}\\[1ex]
\multicolumn{7}{l}{\parbox{0.9\textwidth}{\footnotesize $^b$~MMS designations from \citet{chini97}.}}\\[1ex]
\multicolumn{7}{l}{\parbox{0.9\textwidth}{\footnotesize $^c$~Positions for each flow from Table 3 in \citet*{smz02}.}}\\[1ex]
\multicolumn{7}{l}{\parbox{0.9\textwidth}{\footnotesize $^d$~Jet length from \citet*{smz02} assuming a distance to Orion A of 450 pc.}}\\[1ex]
\multicolumn{7}{l}{\parbox{0.9\textwidth}{\footnotesize $^e$~References: (1) \citet*{rbd97}, (2) \citet{dav07}, (3) \citet*{smz02}, (4) \citet*{ybd97}.}}\\[1ex]

\label{table:hhobjects}
\end{tabular}  }
\end{center}
\end{table}
\setlength{\tabcolsep}{\deftabcolsep}

\subsection{Pre-main Sequence Stars and Infrared Selected Protostars \label{sec:spitzer}}

In the discovery paper of OMC-2, \citet{gbmnps74} presented a
2.2~$\mu$m map of OMC-2 showing five bright sources surrounded by
nebulosity, four without optical counterparts.  Observations from 1.6
to 20~$\mu$m showed that these sources had rising spectral energy
distributions.  \citet{rayner89} published the first infrared array
observations of OMC-2 at 1$-$3~$\mu$m, including polarimetry which
showed that the four infrared sources were the source of the reflection
nebulosity associated with the sources.  Infrared photometry
by \citet{johnson90} found luminosities of 20 to 1000~$L_{\odot}$
for these sources.

\begin{figure}[!ht]
\centering
\includegraphics[width=0.8\textwidth]{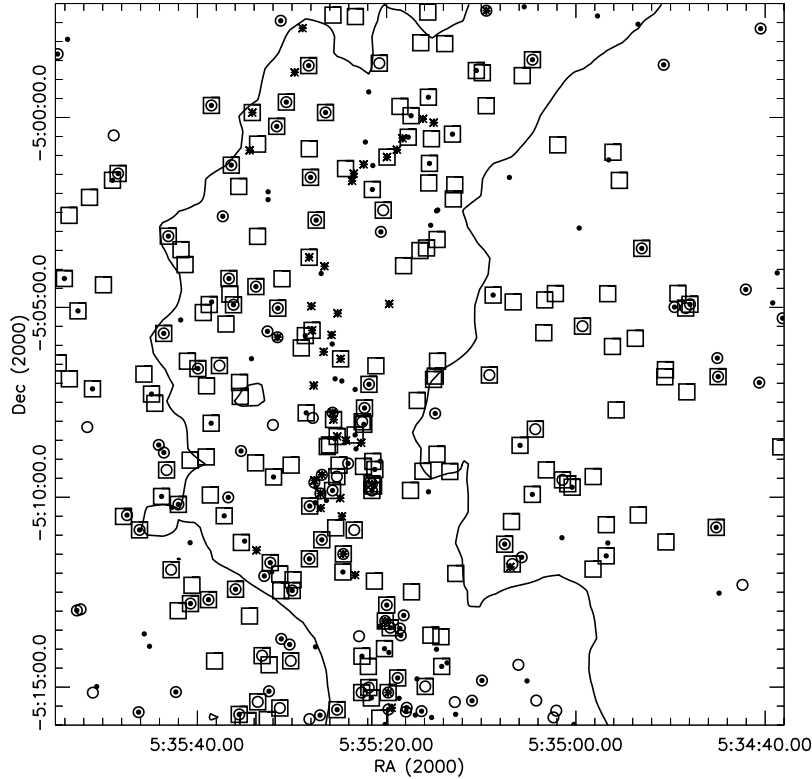}

\caption[Pre-Main Sequence Stars in OMC-2/3]{The pre-main sequence star
population in OMC-2/3 overlaid on the CS $2\rightarrow1$ emission (N. Ridge,
priv. comm.).  The sources represented by dots are those with
infrared excess, asterisks are the infrared-selected protostars, open circles
are the \citet*{chs01} variables, and squares are infrared sources with X-ray
emission detected by Chandra \citep{tsujimoto2002}.}
\label{fig:tom_pms}
\end{figure}

\begin{figure}[!tbp]
\plotfiddle{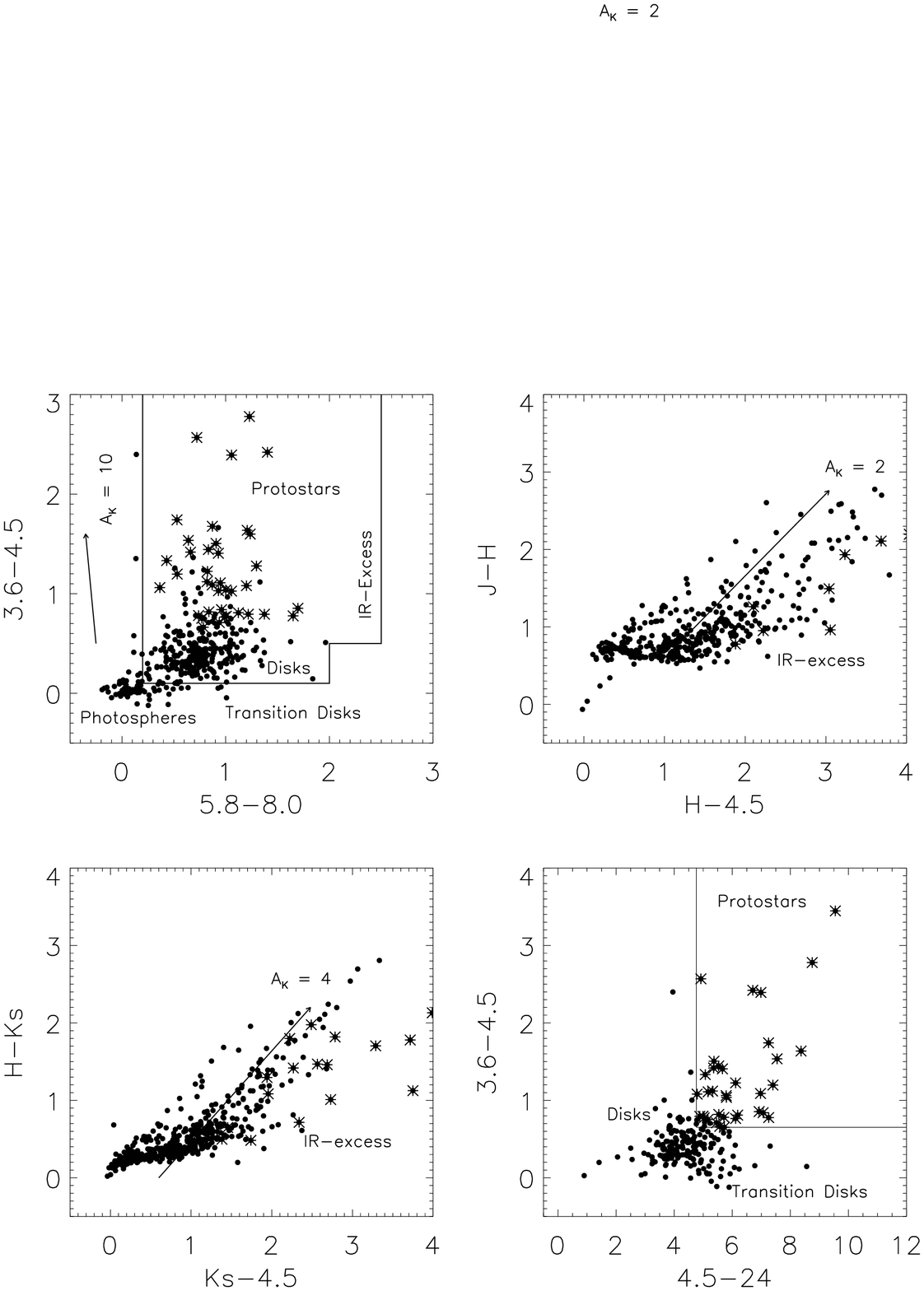}{13.5cm}{0}{69}{69}{-205}{0}
\caption[Mid-Infrared Color-Color Diagrams]{Color-color diagrams of
YSOs in OMC-2/3 and NGC~1977 drawn from 2MASS and {\it Spitzer}
photometry \citep[][in prep]{megeath2007a}.  The YSOs are identified
either by infrared excesses, variability, or, in the case of OMC-2/3,
X-ray emission.  In the upper left diagram, we show the region of the
3.6$-$4.5 versus 5.8$-$8.0 color-color diagram populated by young
stars with disks and protostars
\citep{allen2004,megeath2004,winston2007}.  For sources not detected
at 5.8 and 8.0 $\mu$m, sources with infrared excesses due to disks
and/or envelopes were identified by combining 4.5 $\mu$m photometry
with near-infrared 2MASS $J$, $H$, $K_S$ photometry
\citep{rguter04,winston2007}.  As shown in the upper right and bottom
left diagrams, the infrared excess sources can be separated from
reddened photospheres using the reddening vector from
\citet{flaherty2007}.  Additional transitional disk sources were
identified using the 8.0$-$24 color \citep[][in prep, with modified
criteria]{muz04,winston2007,megeath2007a}; these are grouped with the
stars with disks in our classification scheme.  The bottom right shows
the 3.6$-$4.5 versus 4.5$-$24 color-color diagram used to find likely
protostars by identifying sources with mid-infrared colors consistent
with a flat or rising spectral energy distribution \citep[][in
prep]{megeath2007b}.}
\label{fig:tom_midircc}
\end{figure}

The first wide-field infrared map of OMC-2/3 was a $15' \times 5'$
$JHK$ mosaic by \citet{jones94} which showed OMC-2 as a small group
of red sources surrounded by an extended and relatively uniform
distribution of stars.  They argued that the majority of the 219
sources were associated with the Orion cloud, and were not background
stars; this is the first evidence of a large population of stars
associated with OMC-2.  \citet{ad95} mapped OMC-2 as part of a
2.2~$\mu$m survey of the Orion A molecular cloud.  Their
surface density maps showed OMC-2 as a distinct density peak about
$15'$ north of the center of the ONC; within this density peak they
found $\sim 30$ sources.  They also noted the presence of a large
distributed population surrounding the ONC and OMC-2 regions
containing around $\sim 1500$ members.  \citet*{ncm03} combined mid-infrared
imaging with the 2MASS point source catalog and found 57 sources with
infrared excess in the OMC-2 and OMC-3 regions. Using the 2MASS point
source catalog, they identified 260 possible young stars in the OMC-2
and OMC-3 regions, and found that the absolute $J$-band magnitude distribution
of sources in OMC-2 peaked at a higher luminosity than that for OMC-3.  It is
not clear if this difference is due to a different distribution of stellar
masses or different age distributions; hopefully, future spectroscopic
observations can be used to search for variations in the mass spectrum.

Recent infrared and X-ray observations of the OMC-2/3 region motivate a
re-examination of its population of pre-main sequence stars and protostars.
These include the 2MASS survey of Orion, multi-epoch near-infrared imaging
by \citet*{chs01}, a 100 kilosecond Chandra integration by
\citet{tsujimoto2002}, and finally, a wide field {\it Spitzer} survey of the
Orion molecular clouds \citep[][in prep]{megeath2007a}.  Using catalogs
derived from these observations, we select young stellar objects as sources
satisfying one of the following criteria: 1) presence of a mid-infrared
excess from a dusty disk or envelope
\citep[][in prep; Figure \ref{fig:tom_midircc}]{allen2004,megeath2004,rguter04,winston2007,whitney04,megeath2007a,megeath2007b}, 2)
variability in the near-infrared \citep*{chs01}, or 3) X-ray emission from
an infrared source detected by {\it Spitzer} and/or 2MASS.  We note that deep
near-infrared imaging by \citet{tsujimoto2003} detected 74 X-ray sources which
were not detected by 2MASS; we have not merged our data with their source list.

Using these criteria, we have identified 347 candidate member pre-main
sequence stars and protostars.  Table \ref{table:pms_stars} shows photometry
for the young stellar object candidates that are detected at 24 $\mu$m.  The
X-ray sources are taken from a reprocessing of the Chandra data of
\citet{tsujimoto2002} for the ANCHORS database (An Archive of Chandra
Observations of Regions of Star Formation, S. Wolk P. I.).  In Figure
\ref{fig:tom_pms}, we show the distribution of these sources over the
$20' \times 20'$ field. These sources were also shown in Figure
\ref{fig:tom_proto}.  In Figure \ref{fig:tom_midircc}, we show
color-color diagrams of the sources.  Out of 347 sources total, we identify
208 infrared excess sources.  Of these, 43 sources with 24 $\mu$m detections
are classified as protostars (Class 0/I).  The remaining 165 are predominantly
stars with disks (Class II), although there may also be some protostars
without 24 $\mu$m detections among these stars.  In addition, 123 sources
exhibit variability in the near-infrared, and 197 have X-ray detections.
A total of 139 sources exhibiting variability or X-ray emission do not have
infrared excesses and thus appear to be pure photospheres in the infrared
(Class III).  These observations demonstrate the presence of a large number of
pre-main sequence stars in the OMC-2/3 region.  Future spectroscopic
observations are needed to determine the ages and masses of these stars and
determine the star forming history and initial mass function of the OMC-2/3
region.

\subsection{Young Brown Dwarfs in OMC-2/3}

In addition to a large population of pre-main sequence stars and
protostars, a number of young brown dwarfs have been discovered in OMC-2/3.
Table \ref{table:bdspectra} lists the 19 spectroscopically confirmed young
brown dwarfs from \citet{peterson2007} along with their $I-$ and
$z^{\prime}-$band magnitudes, deep near-infrared $JHK-$band magnitudes, and
spectral types.  The spectral types of these young brown dwarfs range from
M6.5 to M9, which corresponds to approximate masses of 0.075$-$0.015
$M_{\odot}$ using the evolutionary models of \citet{bcah98}.  At least one of
these young brown dwarfs (source 21 in Table \ref{table:bdspectra}) has strong
H$\alpha$ emission, indicating that it is actively accreting.

Young brown dwarfs, with ages much less than 10 Myr, are much more
luminous and thus easier to detect than field brown dwarfs (typically older
than 1 Gyr).  However, the near-infrared colors of young brown dwarfs are
similar to those of very low mass stars, making it very difficult to identify
brown dwarfs from near-infrared photometry alone.  Therefore, techniques
for identifying brown dwarfs in star forming regions \citep{l99, mall01}
were applied by \citet{peterson2007} to select brown dwarf candidates in
OMC-2/3 from near-infrared and visible wavelength photometry.  Once candidate
brown dwarfs were selected, far-red (0.6$-$1.0 $\mu$m) and near-infrared
(0.8$-$2.5 $\mu$m) spectroscopy was obtained and used to confirm the
candidates as {\it bona fide} brown dwarfs.

\begin{figure}[hb]
\centering
\includegraphics[viewport=0 0 380 500,clip=True,width=0.55\textwidth,draft=False]{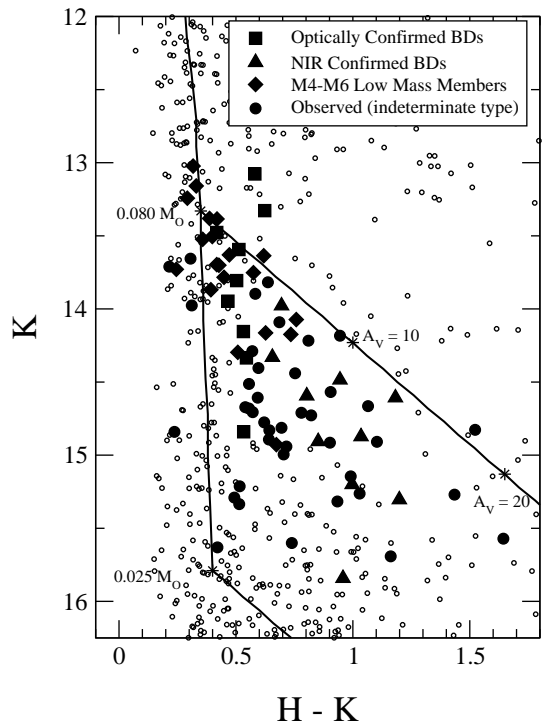}
\caption[Near-Infrared Selection of Brown Dwarfs]{$K$ vs. $H-K$
color-magnitude diagram of all the stars detected at $H$ and $K$ in
OMC-2/3 (small open circles) from \citet{peterson2007}.  Over-plotted is the
\citet{bcah98} evolutionary isochrone for a 1 Myr-old pre-main sequence star,
assuming a distance to OMC-2/3 of 450 pc.  Reddening vectors are plotted for:
1) a 0.08 $M_{\odot}$ star (upper diagonal line) and 2) a 25 $M_J$ brown dwarf
with an age of 1 Myr (lower diagonal line).  The sources confirmed as brown
dwarfs (spectral types of M6.5$-$M9) from far-red spectroscopy are represented
by squares; those confirmed as low mass members (spectral types of M4$-$M6)
are represented by diamonds.  The sources confirmed as brown dwarfs from
near-infrared spectroscopy are represented by triangles.  All sources whose
spectral types could not be determined or sources which were determined to be
foreground or background dwarfs are represented by filled circles.  Tables of
photometry for all the sources whose spectra were obtained can be found in
\citet{peterson2007}.}
\label{fig:k_hk}
\end{figure}

Figure \ref{fig:k_hk} is a $K$ versus $H-K$ color-magnitude diagram of
all sources in OMC-2/3 from \citet{peterson2007}.  Assuming an age of 1~Myr,
\citet{bcah98} isochrones were used to select sources below the hydrogen
burning limit as candidate brown dwarfs.  For sources that were not as deeply
embedded and detected in the $I-$ and $z^{\prime}-$bands, near-infrared plus
visible wavelength photometry was used for selection.  The sources
confirmed by spectroscopy as brown dwarfs or low-mass stellar members are
shown in Figure \ref{fig:k_hk}.  Note that we have observed all brown
dwarf candidates that fall below 0.080 $M_{\odot}$ and above 0.050 $M_{\odot}$
(for 1~Myr objects) with an extinction limit of $4.5 < A_V < 14$, making a
complete sample within this regime.  Figure \ref{fig:k_hk} illustrates how
spectroscopy is needed to confirm the mass of the object, with both young
($<$ 1~Myr) brown dwarfs found above the reddening vector for 1~Myr old
0.08 $M_{\odot}$ and older low-mass stellar members found below the vector.
Interpretations for this apparent dispersion in near-infrared luminosities are
discussed in \citet{peterson2007}.

\setlength{\tabcolsep}{1.4\deftabcolsep}

\begin{landscape}
{\scriptsize

\begin{longtable}{cccccccccccc}

\caption{Young Stellar Object Candidates in OMC-2/3 detected at 24 $\mu$m}\\
\noalign{\smallskip}
\hline
\noalign{\smallskip}
  RA &	Dec &	K$_S$\tablenotemark{a} & 3.6~$\mu$m & 4.5~$\mu$m & 5.8~$\mu$m & 8~$\mu$m & 24~$\mu$m & Var. & X-ray & IR-ex\tablenotemark{b} & Proto.\\
  J(2000) & J(2000)  &	mag (unc) & mag (unc) & mag (unc) & mag (unc) & mag (unc) & mag (unc)  & & & &\\
\noalign{\smallskip}
\hline
\noalign{\smallskip}
\endfirsthead

\caption{Young Stellar Object Candidates in OMC-2/3 detected at 24~$\mu$m (continued)}\\
\noalign{\smallskip}
\hline
\noalign{\smallskip}
  RA &	Dec &	K$_S$\tablenotemark{a} & 3.6~$\mu$m & 4.5~$\mu$m & 5.8~$\mu$m & 8~$\mu$m & 24~$\mu$m & Var. & X-ray & IR-ex\tablenotemark{b} & Proto.\\
  J(2000) & J(2000)  &	mag (unc) & mag (unc) & mag (unc) & mag (unc) & mag (unc) & mag (unc)  & & & &\\
\noalign{\smallskip}
\hline
\noalign{\smallskip}
\endhead

\noalign{\smallskip}
\hline
\noalign{\smallskip}
\multicolumn{12}{l}{{$^a$ Magnitudes from 2MASS photometry.}}\\[1ex]
\multicolumn{12}{l}{{$^b$ From \citet*{chs01}; Those that are variable have Stetson indices greater than 0.55.}}\\[1ex]
\endfoot

\noalign{\smallskip}
\hline
\noalign{\smallskip}
\multicolumn{12}{l}{{$^a$ Magnitudes from 2MASS photometry.}}\\[1ex]
\multicolumn{12}{l}{{$^b$ From \citet*{chs01}; Those that are variable have Stetson indices greater than 0.55.}}\\[1ex]
\endlastfoot

 5 35 19.5 & -5 15 32 &   &   &  9.55 (0.01) &  8.12 (0.01) &  7.16 (0.03) &  1.64 (0.03) & 0 &  0 &  0 &  1 \\
 5 35 11.0 & -5 15 21 &  11.12 (0.02) &  9.23 (0.01) &  8.46 (0.01) &  8.10 (0.01) &  7.71 (0.03) &  4.32 (0.07) & 1 &  0 &  1 &  0 \\
 5 35 19.8 & -5 15 08 &  10.21 (0.03) &  8.56 (0.01) &  7.52 (0.01) &  6.67 (0.01) &  5.67 (0.02) &  1.72 (0.03) & 1 &  1 &  0 &  1 \\
 5 35 05.2 & -5 14 50 &   7.19 (0.02) &  6.99 (0.01) &  6.67 (0.01) &  6.39 (0.01) &  5.97 (0.01) &  3.71 (0.02) & 0 &  0 &  1 &  0 \\
 5 35 20.2 & -5 13 59 &   9.80 (0.02) &  8.97 (0.01) &  8.53 (0.01) &  8.01 (0.01) &  7.28 (0.03) &  3.24 (0.02) & 0 &  1 &  1 &  0 \\
 5 35 18.5 & -5 13 38 &   9.29 (0.02) &  7.73 (0.01) &  7.03 (0.01) &  6.16 (0.01) &  5.27 (0.01) &  1.79 (0.01) & 1 &  0 &  1 &  0 \\
 5 35 45.6 & -5 13 35 &  11.67 (0.02) & 10.67 (0.01) & 10.30 (0.01) &  9.97 (0.01) &  9.54 (0.07) &  4.66 (0.05) & 0 &  0 &  1 &  0 \\
 5 35 20.1 & -5 13 15 &   6.43 (0.03) &  6.16 (0.01) &  4.48 (0.01) &  3.67 (0.01) &  2.80 (0.01) & -0.38 (0.02) & 1 &  1 &  0 &  1 \\
 5 35 18.2 & -5 13 06 &   9.64 (0.02) &  8.67 (0.01) &  8.19 (0.01) &  7.83 (0.01) &  7.16 (0.02) &  3.84 (0.07) & 1 &  0 &  1 &  0 \\
 5 35 52.7 & -5 12 58 &  10.32 (0.02) &  9.63 (0.01) &  9.31 (0.01) &  9.03 (0.01) &  8.25 (0.01) &  4.76 (0.06) & 1 &  0 &  1 &  0 \\
 5 35 38.8 & -5 12 41 &   8.82 (0.02) &  7.97 (0.01) &  7.57 (0.01) &  6.96 (0.01) &  6.25 (0.02) &  3.72 (0.05) & 1 &  1 &  1 &  0 \\
 5 34 44.9 & -5 12 31 &  13.82 (0.04) & 13.09 (0.02) & 12.74 (0.02) & 12.35 (0.10) &   &  8.28 (0.02) & 0 &  0 &  1 &  0 \\
 5 35 36.0 & -5 12 25 &   8.39 (0.02) &  7.17 (0.01) &  6.52 (0.01) &  6.04 (0.01) &  5.30 (0.02) &  2.70 (0.01) & 1 &  1 &  1 &  0 \\
 5 35 32.9 & -5 12 04 &  10.29 (0.02) &  9.12 (0.01) &  8.66 (0.01) &  8.32 (0.01) &  7.84 (0.03) &  4.75 (0.04) & 1 &  0 &  1 &  0 \\
 5 35 23.3 & -5 12 03 &    & 10.49 (0.01) &  8.96 (0.01) &  8.58 (0.04) &  7.94 (0.08) &  1.41 (0.02) & 0 &  0 &  0 &  1 \\
 5 35 06.9 & -5 11 50 &    &   & 13.70 (0.04) &   &   &  5.31 (0.07) & 0 &  0 &  0 &  1 \\
 5 35 28.2 & -5 11 37 &  11.32 (0.02) & 10.54 (0.01) & 10.10 (0.01) &  9.65 (0.03) &   &  4.74 (0.11) & 1 &  1 &  1 &  0 \\
 5 35 05.7 & -5 11 34 &  10.76 (0.02) & 10.17 (0.01) &  9.64 (0.01) &  9.32 (0.01) &  8.57 (0.04) &  5.22 (0.04) & 1 &  0 &  1 &  0 \\
 5 34 56.8 & -5 11 32 &  10.15 (0.02) &  8.87 (0.01) &  8.57 (0.01) &  8.19 (0.01) &  7.45 (0.01) &  4.55 (0.02) & 0 &  1 &  1 &  0 \\
 5 35 24.6 & -5 11 29 &  10.39 (0.02) &  9.64 (0.01) &  8.93 (0.01) &  8.51 (0.01) &  7.57 (0.03) &  3.47 (0.02) & 1 &  1 &  1 &  0 \\
 5 35 33.8 & -5 11 23 &    &   & 13.30 (0.07) &   &   &  5.30 (0.05) & 0 &  0 &  0 &  1 \\
 5 35 07.5 & -5 11 14 &  10.83 (0.02) & 10.29 (0.01) & 10.09 (0.01) &  9.89 (0.01) &  9.17 (0.05) &  5.18 (0.09) & 1 &  1 &  1 &  0 \\
 5 34 56.6 & -5 11 12 &  12.20 (0.02) & 11.44 (0.01) & 10.94 (0.01) & 10.58 (0.03) &  9.86 (0.06) &  6.31 (0.05) & 0 &  0 &  1 &  0 \\
 5 35 26.8 & -5 11 07 &   7.02 (0.03) &  5.80 (0.01) &  5.21 (0.01) &  4.78 (0.01) &  4.12 (0.01) &  1.82 (0.08) & 1 &  1 &  1 &  0 \\
 5 35 01.5 & -5 11 03 &  12.23 (0.03) & 11.63 (0.01) & 11.20 (0.01) & 10.81 (0.04) &  9.54 (0.07) &  6.27 (0.01) & 0 &  0 &  1 &  0 \\
 5 35 46.1 & -5 10 51 &  10.43 (0.02) &  9.86 (0.01) &  9.46 (0.01) &  9.43 (0.01) &  8.75 (0.03) &  5.18 (0.01) & 1 &  1 &  1 &  0 \\
 5 34 45.2 & -5 10 47 &  10.70 (0.02) & 10.15 (0.01) &  9.77 (0.01) &  9.63 (0.01) &  9.13 (0.02) &  6.11 (0.06) & 1 &  1 &  1 &  0 \\
 5 35 24.7 & -5 10 30 &   9.59 (0.07) &  7.32 (0.01) &  5.73 (0.01) &  4.80 (0.01) &  3.56 (0.01) & -1.40 (0.03) & 0 &  0 &  0 &  1 \\
 5 35 47.4 & -5 10 28 &   9.77 (0.03) &  8.49 (0.01) &  8.02 (0.01) &  7.64 (0.01) &  6.93 (0.01) &  4.25 (0.02) & 1 &  0 &  1 &  0 \\
 5 35 27.0 & -5 10 17 &   8.80 (0.03) &  5.83 (0.01) &  4.81 (0.01) &  3.69 (0.01) &  2.63 (0.01) & -0.70 (0.50) & 0 &  0 &  1 &  0 \\
 5 35 42.0 & -5 10 11 &   9.47 (0.02) &  8.73 (0.01) &  8.42 (0.01) &  8.00 (0.01) &  7.42 (0.01) &  4.56 (0.11) & 1 &  1 &  1 &  0 \\
 5 35 24.9 & -5 10 01 &  11.77 (0.19) & 10.13 (0.05) &  8.71 (0.03) &  7.90 (0.02) &  7.24 (0.03) &  3.36 (0.01) & 0 &  0 &  0 &  1 \\
 5 35 36.7 & -5 10 00 &  11.94 (0.02) & 11.04 (0.01) & 10.53 (0.01) & 10.11 (0.03) &  9.25 (0.06) &  6.17 (0.03) & 1 &  0 &  1 &  0 \\
 5 35 04.6 & -5 09 55 &  10.39 (0.02) & 10.21 (0.01) & 10.12 (0.01) & 10.06 (0.01) & 10.09 (0.07) &  5.28 (0.02) & 0 &  1 &  1 &  0 \\
 5 35 27.0 & -5 09 54 &  12.75 (0.02) & 10.23 (0.02) &  9.00 (0.04) &  8.13 (0.03) &  7.31 (0.02) &  2.89 (0.03) & 1 &  0 &  0 &  1 \\
 5 35 15.6 & -5 09 51 &  11.54 (0.02) & 10.55 (0.01) & 10.04 (0.01) &  9.57 (0.01) &  8.92 (0.03) &  5.57 (0.03) & 0 &  0 &  1 &  0 \\
 5 35 25.7 & -5 09 49 &  10.11 (0.02) &  9.61 (0.01) &  9.56 (0.03) &  9.27 (0.05) &  8.34 (0.06) &  3.63 (0.01) & 1 &  1 &  1 &  0 \\
 5 35 21.6 & -5 09 49 &  11.67 (0.02) & 10.80 (0.01) & 10.40 (0.01) & 10.06 (0.02) &  9.18 (0.04) &  5.73 (0.06) & 1 &  1 &  1 &  0 \\
 5 35 00.4 & -5 09 44 &  12.78 (0.03) & 12.42 (0.01) & 12.26 (0.01) & 11.87 (0.05) & 10.98 (0.07) &  5.49 (0.06) & 0 &  1 &  1 &  0 \\
 5 35 21.6 & -5 09 38 &  12.44 (0.02) & 10.86 (0.01) & 10.10 (0.01) &  9.38 (0.01) &  8.39 (0.01) &  3.98 (0.05) & 1 &  1 &  0 &  1 \\
 5 35 31.9 & -5 09 28 &   8.03 (0.02) &  7.29 (0.01) &  6.83 (0.01) &  6.40 (0.01) &  5.49 (0.01) &  2.37 (0.04) & 0 &  1 &  1 &  0 \\
 5 35 26.8 & -5 09 24 &  11.15 (0.05) &  6.81 (0.01) &  5.78 (0.01) &  4.27 (0.01) &  3.34 (0.01) &  0.65 (0.25) & 1 &  0 &  1 &  0 \\
 5 35 21.2 & -5 09 16 &   7.10 (0.03) &  6.28 (0.01) &  6.01 (0.01) &  5.88 (0.01) &  5.54 (0.01) &  3.96 (0.02) & 0 &  1 &  1 &  0 \\
 5 35 23.2 & -5 08 43 &  12.98 (0.02) & 11.14 (0.01) & 10.69 (0.01) &  9.91 (0.02) &  9.17 (0.04) &  5.78 (0.03) & 0 &  0 &  1 &  0 \\
 5 35 24.3 & -5 08 30 &    & 11.33 (0.01) &  8.91 (0.01) &  7.32 (0.01) &  5.92 (0.01) &  2.20 (0.04) & 0 &  0 &  0 &  1 \\
 5 35 25.2 & -5 08 23 &    & 12.71 (0.02) & 10.14 (0.01) &  8.86 (0.01) &  8.13 (0.02) &  5.22 (0.03) & 0 &  1 &  0 &  1 \\
 5 35 23.3 & -5 08 21 &  12.49 (0.03) & 10.57 (0.01) &  9.79 (0.01) &  9.22 (0.01) &  8.51 (0.02) &  5.61 (0.03) & 0 &  1 &  1 &  0 \\
 5 35 04.3 & -5 08 12 &   7.27 (0.02) &  7.15 (0.01) &  7.12 (0.01) &  7.09 (0.01) &  6.97 (0.01) &  6.22 (0.03) & 1 &  1 &  0 &  0 \\
 5 35 22.4 & -5 08 04 &   8.61 (0.04) &  7.25 (0.01) &  6.66 (0.01) &  6.08 (0.01) &  4.98 (0.01) &  1.50 (0.06) & 0 &  1 &  1 &  0 \\
 5 35 38.5 & -5 08 03 &  12.01 (0.02) & 10.82 (0.01) & 10.25 (0.01) & 10.14 (0.01) &  9.70 (0.03) &  6.24 (0.31) & 0 &  1 &  1 &  0 \\
 5 35 22.6 & -5 08 00 &   9.21 (0.03) &  8.82 (0.01) &  8.68 (0.01) &  8.26 (0.01) &   &  2.61 (0.04) & 1 &  1 &  0 &  0 \\
 5 35 25.6 & -5 07 57 &  10.27 (0.06) &  9.36 (0.01) &  8.52 (0.01) &  7.73 (0.01) &  6.77 (0.02) &  1.46 (0.02) & 0 &  1 &  0 &  1 \\
 5 35 25.7 & -5 07 46 &  11.23 (0.03) &  9.62 (0.01) &  8.96 (0.01) &  8.40 (0.01) &  7.61 (0.03) &  3.39 (0.04) & 1 &  0 &  1 &  0 \\
 5 35 22.3 & -5 07 38 &  10.26 (0.03) &  9.04 (0.01) &  8.32 (0.01) &  7.56 (0.01) &  6.54 (0.02) &  3.28 (0.02) & 1 &  1 &  1 &  0 \\
 5 35 44.8 & -5 07 17 &   8.91 (0.02) &  8.34 (0.01) &  8.27 (0.01) &  7.86 (0.01) &  6.90 (0.01) &  3.16 (0.03) & 0 &  1 &  1 &  0 \\
 5 35 51.1 & -5 07 08 &   9.87 (0.02) &  9.49 (0.01) &  9.20 (0.01) &  8.85 (0.01) &  7.85 (0.01) &  4.84 (0.01) & 0 &  1 &  1 &  0 \\
 5 35 27.7 & -5 07 03 &    &   & 12.61 (0.06) &   &   &  3.96 (0.06) & 0 &  0 &  0 &  1 \\
 5 35 21.9 & -5 07 01 &   9.32 (0.02) &  8.38 (0.01) &  7.97 (0.01) &  7.72 (0.01) &  6.94 (0.03) &  3.21 (0.03) & 1 &  1 &  1 &  0 \\
 5 34 40.6 & -5 06 58 &  11.82 (0.02) & 10.88 (0.01) & 10.36 (0.01) & 10.02 (0.01) &  9.47 (0.01) &  6.32 (0.03) & 1 &  0 &  1 &  0 \\
 5 34 45.0 & -5 06 49 &  10.66 (0.02) &  9.94 (0.01) &  9.79 (0.01) &  9.35 (0.01) &  8.43 (0.01) &  5.22 (0.06) & 1 &  1 &  1 &  0 \\
 5 35 39.9 & -5 06 36 &  10.80 (0.02) & 10.05 (0.01) &  9.60 (0.01) &  9.20 (0.01) &  8.39 (0.01) &  5.55 (0.05) & 1 &  1 &  1 &  0 \\
 5 35 34.3 & -5 06 21 &   9.32 (0.02) &  9.36 (0.01) &  9.35 (0.01) &  9.38 (0.01) &  9.32 (0.04) &  5.65 (0.04) & 0 &  0 &  1 &  0 \\
 5 35 24.9 & -5 06 21 &  10.79 (0.02) &  9.09 (0.01) &  8.30 (0.01) &  7.70 (0.01) &  6.96 (0.01) &  2.59 (0.03) & 0 &  1 &  0 &  1 \\
 5 34 45.1 & -5 06 20 &  12.23 (0.02) & 11.56 (0.01) & 11.19 (0.01) & 10.88 (0.02) & 10.34 (0.04) &  7.25 (0.01) & 1 &  0 &  1 &  0 \\
 5 35 26.7 & -5 06 10 &    & 12.26 (0.01) &  9.87 (0.01) &  8.34 (0.01) &  7.28 (0.01) &  2.88 (0.07) & 0 &  0 &  0 &  1 \\
 5 35 25.8 & -5 05 57 &  12.65 (0.05) & 10.04 (0.01) &  8.68 (0.01) &  7.85 (0.01) &  7.16 (0.01) &  4.11 (0.01) & 0 &  0 &  1 &  0 \\
 5 35 31.5 & -5 05 47 &   8.26 (0.03) &  6.47 (0.01) &  5.70 (0.01) &  4.99 (0.01) &  3.98 (0.01) &  0.66 (0.01) & 1 &  0 &  0 &  1 \\
 5 35 28.6 & -5 05 44 &  10.80 (0.02) &  9.24 (0.01) &  8.66 (0.01) &  8.13 (0.01) &  7.58 (0.01) &  4.34 (0.01) & 0 &  1 &  1 &  0 \\
 5 35 25.8 & -5 05 43 &    &   & 11.85 (0.03) & 11.96 (0.09) &   &  4.25 (0.04) & 0 &  0 &  0 &  1 \\
 5 35 43.6 & -5 05 41 &  10.26 (0.02) &  9.31 (0.01) &  8.59 (0.01) &  8.22 (0.01) &  7.35 (0.01) &  4.38 (0.13) & 1 &  1 &  1 &  0 \\
 5 35 27.9 & -5 05 36 &  13.50 (0.04) & 11.30 (0.01) & 10.55 (0.01) &  9.96 (0.02) &  9.19 (0.04) &  5.39 (0.05) & 0 &  1 &  0 &  1 \\
 5 35 25.2 & -5 05 09 &    & 12.60 (0.01) & 11.09 (0.01) & 10.17 (0.01) &  9.26 (0.02) &  5.73 (0.02) & 0 &  0 &  0 &  1 \\
 5 35 52.6 & -5 05 05 &   8.75 (0.02) &  7.99 (0.01) &  7.73 (0.01) &  7.52 (0.01) &  7.03 (0.01) &  3.61 (0.01) & 0 &  1 &  1 &  0 \\
 5 35 31.5 & -5 05 01 &  10.82 (0.02) &  9.89 (0.01) &  9.14 (0.01) &  8.98 (0.01) &  7.93 (0.01) &  4.86 (0.04) & 1 &  1 &  1 &  0 \\
 5 34 49.6 & -5 04 59 &  12.25 (0.02) & 11.86 (0.01) & 11.61 (0.01) & 11.40 (0.02) & 10.87 (0.02) &  7.94 (0.01) & 1 &  0 &  1 &  0 \\
 5 35 28.0 & -5 04 58 &    &   & 12.20 (0.02) & 11.88 (0.05) &   &  5.29 (0.08) & 0 &  0 &  0 &  1 \\
 5 35 36.2 & -5 04 55 &  10.99 (0.02) & 10.28 (0.01) &  9.96 (0.01) &  9.78 (0.01) &  9.16 (0.02) &  5.25 (0.01) & 1 &  1 &  1 &  0 \\
 5 35 19.7 & -5 04 54 &  14.03 (0.04) & 10.41 (0.01) &  9.01 (0.01) &  7.92 (0.01) &  6.99 (0.01) &  3.35 (0.03) & 0 &  0 &  0 &  1 \\
 5 34 47.9 & -5 04 55 &  10.62 (0.02) & 10.05 (0.01) &  9.81 (0.01) &  9.64 (0.01) &  8.85 (0.01) &  4.88 (0.03) & 1 &  1 &  1 &  0 \\
 5 35 38.5 & -5 04 51 &  12.16 (0.02) & 10.47 (0.01) &  9.77 (0.01) &  9.31 (0.01) &  8.29 (0.01) &  5.01 (0.02) & 0 &  0 &  1 &  0 \\
 5 35 08.7 & -5 04 40 &  11.12 (0.02) & 10.75 (0.01) & 10.37 (0.01) & 10.15 (0.01) &  9.42 (0.01) &  6.76 (0.02) & 0 &  1 &  1 &  0 \\
 5 34 42.0 & -5 04 31 &  12.61 (0.02) & 11.99 (0.01) & 11.52 (0.01) & 11.22 (0.02) & 10.51 (0.02) &  7.80 (0.05) & 1 &  0 &  1 &  0 \\
 5 35 33.8 & -5 04 27 &  10.46 (0.02) &  9.11 (0.01) &  8.52 (0.01) &  8.11 (0.01) &  7.44 (0.01) &  3.87 (0.02) & 1 &  1 &  1 &  0 \\
 5 35 54.1 & -5 04 14 &   9.83 (0.03) &  9.09 (0.01) &  8.71 (0.01) &  8.53 (0.01) &  8.01 (0.01) &  6.20 (0.06) & 0 &  1 &  1 &  0 \\
 5 35 36.7 & -5 04 14 &  10.77 (0.02) & 10.24 (0.01) &  9.96 (0.01) &  9.76 (0.01) &  9.28 (0.02) &  5.51 (0.01) & 1 &  1 &  1 &  0 \\
 5 35 26.6 & -5 03 55 &    &  8.35 (0.01) &  7.07 (0.01) &  5.91 (0.01) &  4.61 (0.01) & -0.94 (0.02) & 0 &  0 &  0 &  1 \\
 5 35 18.2 & -5 03 54 &   9.15 (0.02) &  9.19 (0.01) &  9.15 (0.01) &  9.14 (0.01) &  8.97 (0.02) &  6.28 (0.03) & 0 &  1 &  0 &  0 \\
 5 35 28.2 & -5 03 40 &  12.50 (0.02) &  7.47 (0.01) &  6.41 (0.01) &  5.11 (0.01) &  4.74 (0.01) &  0.62 (0.01) & 0 &  1 &  0 &  1 \\
 5 34 53.0 & -5 03 26 &  10.79 (0.02) & 10.18 (0.01) &  9.80 (0.01) &  9.40 (0.01) &  8.65 (0.01) &  5.91 (0.05) & 1 &  1 &  1 &  0 \\
 5 35 43.1 & -5 03 07 &  12.28 (0.02) & 11.88 (0.01) & 11.60 (0.01) & 11.27 (0.02) & 10.68 (0.04) &  7.32 (0.35) & 1 &  1 &  1 &  0 \\
 5 35 20.6 & -5 03 00 &  11.83 (0.03) &  9.52 (0.01) &  8.85 (0.01) &  8.25 (0.01) &  7.34 (0.01) &  3.81 (0.03) & 1 &  0 &  1 &  0 \\
 5 35 15.3 & -5 02 50 &  14.19 (0.05) & 12.91 (0.02) & 12.43 (0.02) & 11.88 (0.04) & 11.00 (0.05) &  8.34 (0.04) & 0 &  0 &  1 &  0 \\
 5 35 27.4 & -5 02 42 &  10.98 (0.02) &  9.28 (0.01) &  8.66 (0.01) &  8.18 (0.01) &  7.65 (0.01) &  4.83 (0.03) & 1 &  1 &  1 &  0 \\
 5 35 37.3 & -5 02 36 &  12.07 (0.02) & 10.76 (0.01) & 10.18 (0.01) & 10.03 (0.01) &  9.47 (0.02) &  6.28 (0.01) & 1 &  0 &  1 &  0 \\
 5 35 14.7 & -5 02 27 &  11.55 (0.05) & 10.63 (0.01) & 10.39 (0.01) & 10.06 (0.01) &  9.80 (0.02) &  6.67 (0.06) & 0 &  0 &  1 &  0 \\
 5 35 32.5 & -5 01 57 &  12.75 (0.03) & 11.92 (0.01) & 11.58 (0.01) & 10.93 (0.03) &  9.81 (0.05) &  6.93 (0.01) & 0 &  0 &  1 &  0 \\
 5 35 23.6 & -5 01 40 &  15.50 (0.15) & 11.76 (0.01) & 10.12 (0.01) &  8.20 (0.01) &  6.99 (0.01) &  1.75 (0.01) & 0 &  0 &  0 &  1 \\
 5 35 28.0 & -5 01 34 &  12.50 (0.02) & 10.23 (0.01) &  9.43 (0.01) &  8.94 (0.01) &  8.32 (0.01) &  5.58 (0.01) & 1 &  1 &  1 &  0 \\
 5 35 48.4 & -5 01 28 &  10.53 (0.02) &  9.50 (0.01) &  9.05 (0.01) &  8.81 (0.01) &  8.11 (0.01) &  4.35 (0.01) & 1 &  1 &  1 &  0 \\
 5 35 23.5 & -5 01 28 &    & 13.97 (0.02) & 11.19 (0.01) & 10.42 (0.01) &  9.20 (0.02) &  2.44 (0.20) & 0 &  0 &  0 &  1 \\
 5 35 21.5 & -5 01 15 &  13.28 (0.08) & 12.46 (0.01) & 10.06 (0.01) & 10.63 (0.02) & 10.49 (0.09) &  6.11 (0.03) & 0 &  0 &  1 &  0 \\
 5 35 36.4 & -5 01 15 &   9.13 (0.02) &  8.59 (0.01) &  8.32 (0.01) &  8.02 (0.01) &  7.44 (0.01) &  5.09 (0.02) & 1 &  1 &  1 &  0 \\
 5 35 22.4 & -5 01 14 &    & 16.22 (0.09) & 12.77 (0.02) & 12.04 (0.04) & 11.88 (0.11) &  3.23 (0.05) & 0 &  0 &  0 &  1 \\
 5 34 56.5 & -5 01 06 &  13.05 (0.03) & 12.66 (0.01) & 12.22 (0.01) & 11.72 (0.03) & 10.77 (0.04) &  7.38 (0.06) & 0 &  0 &  1 &  0 \\
 5 35 20.0 & -5 01 02 &  14.47 (0.06) & 10.16 (0.01) &  8.82 (0.01) &  7.78 (0.01) &  7.35 (0.01) &  3.75 (0.02) & 0 &  1 &  0 &  1 \\
 5 35 34.5 & -5 00 52 &  11.48 (0.03) &  8.24 (0.01) &  7.12 (0.01) &  6.29 (0.01) &  5.47 (0.01) &  1.80 (0.04) & 0 &  0 &  0 &  1 \\
 5 35 18.9 & -5 00 50 &    &   & 12.58 (0.07) & 10.16 (0.03) &  8.76 (0.01) &  4.06 (0.01) & 0 &  0 &  0 &  1 \\
 5 35 18.3 & -5 00 32 &  11.07 (0.04) &  7.41 (0.01) &  5.97 (0.01) &  5.07 (0.01) &  4.24 (0.01) &  0.44 (0.01) & 0 &  0 &  0 &  1 \\
 5 35 13.0 & -5 00 26 &  14.81 (0.08) & 13.11 (0.01) & 12.53 (0.02) & 12.16 (0.05) & 12.05 (0.09) &  6.83 (0.01) & 0 &  1 &  1 &  0 \\
 5 35 31.6 & -5 00 14 &  10.13 (0.02) &  8.17 (0.01) &  7.71 (0.01) &  7.19 (0.01) &  6.78 (0.01) &  4.28 (0.03) & 1 &  1 &  1 &  0 \\
 5 35 15.0 & -5 00 08 &  14.66 (0.14) & 12.86 (0.01) & 11.66 (0.01) & 11.03 (0.03) & 10.50 (0.04) &  4.26 (0.02) & 0 &  0 &  0 &  1 \\
 5 35 16.2 & -5 00 02 &  11.98 (0.04) &  9.35 (0.01) &  8.26 (0.01) &  7.42 (0.01) &  6.55 (0.01) &  1.29 (0.01) & 0 &  0 &  0 &  1 \\
 5 35 34.2 & -4 59 52 &    & 14.13 (0.04) & 12.39 (0.02) & 11.13 (0.04) & 10.60 (0.08) &  5.14 (0.02) & 0 &  1 &  0 &  1 \\
 5 35 26.5 & -4 59 52 &  12.90 (0.03) & 11.35 (0.01) & 11.05 (0.01) & 10.38 (0.02) &  9.77 (0.03) &  6.78 (0.03) & 1 &  1 &  1 &  0 \\
 5 35 38.5 & -4 59 41 &  10.35 (0.02) &  9.77 (0.01) &  9.39 (0.01) &  9.17 (0.01) &  8.38 (0.02) &  5.50 (0.01) & 1 &  1 &  1 &  0 \\
 5 35 30.6 & -4 59 35 &  10.35 (0.02) &  9.06 (0.01) &  8.50 (0.01) &  8.14 (0.01) &  7.36 (0.01) &  3.62 (0.05) & 1 &  1 &  1 &  0 \\
 5 35 15.6 & -4 59 27 &  11.51 (0.02) & 10.29 (0.01) &  9.77 (0.01) &  9.16 (0.04) &  8.12 (0.07) &  4.39 (0.05) & 0 &  1 &  1 &  0 \\
 5 35 29.7 & -4 58 48 &    &   & 13.73 (0.06) & 11.00 (0.03) &  9.55 (0.03) &  4.21 (0.07) & 0 &  0 &  0 &  1 \\
 5 35 10.5 & -4 58 45 &  10.57 (0.02) &  7.24 (0.01) &  6.24 (0.01) &  5.57 (0.01) &  4.98 (0.01) &  2.59 (0.14) & 0 &  1 &  1 &  0 \\
 5 35 28.2 & -4 58 38 &  11.57 (0.03) &  9.13 (0.01) &  8.24 (0.01) &  7.46 (0.01) &  6.85 (0.01) &  4.88 (0.06) & 1 &  1 &  1 &  0 \\
 5 34 50.7 & -4 58 36 &  10.74 (0.02) & 10.15 (0.01) &  9.92 (0.01) &  9.66 (0.01) &  9.21 (0.01) &  7.37 (0.01) & 1 &  0 &  1 &  0 \\
 5 35 04.6 & -4 58 28 &  11.09 (0.02) &  9.97 (0.01) &  9.62 (0.01) &  9.33 (0.01) &  8.68 (0.01) &  5.53 (0.04) & 1 &  1 &  1 &  0 \\
 5 35 54.8 & -4 58 19 &  12.66 (0.03) & 11.75 (0.01) & 11.26 (0.01) & 10.62 (0.02) &  9.55 (0.02) &  5.75 (0.08) & 1 &  0 &  1 &  0 \\
 5 34 40.5 & -4 57 39 &  11.75 (0.02) & 11.21 (0.01) & 10.95 (0.01) & 10.86 (0.02) & 10.19 (0.02) &  6.40 (0.01) & 1 &  0 &  1 &  0 \\
 5 34 53.4 & -4 57 32 &  12.27 (0.02) & 11.83 (0.01) & 11.49 (0.01) & 11.10 (0.02) & 10.45 (0.03) &  7.70 (0.03) & 0 &  0 &  1 &  0 \\
 5 34 57.7 & -4 57 19 &  13.02 (0.03) & 11.40 (0.01) & 10.78 (0.01) & 10.28 (0.02) &  9.72 (0.03) &  5.79 (0.05) & 0 &  0 &  1 &  0 \\
 5 35 09.5 & -4 57 11 &  12.71 (0.02) & 12.12 (0.01) & 11.33 (0.01) & 10.83 (0.05) &  9.61 (0.08) &  6.34 (0.02) & 1 &  0 &  0 &  1 \\
\label{table:pms_stars}
\end{longtable}
}
\end{landscape}
\setlength{\tabcolsep}{\deftabcolsep}


\begin{landscape}
\begin{table}[!ht]
\caption{Young Brown Dwarfs in OMC-2/3\tablenotemark{k}}
\smallskip
{
\scriptsize
\begin{tabular}{lcccccclcc}
\tableline
\noalign{\smallskip}
 Name & RA\tablenotemark{a} & Dec.\tablenotemark{a} & I, ${\sigma}_I$\tablenotemark{b} &
J, ${\sigma}_J$\tablenotemark{c} & H, ${\sigma}_H$\tablenotemark{c} &
K, ${\sigma}_K$\tablenotemark{c} & Spectral Type&A$_H$\tablenotemark{i}&EW[H$\alpha$]\\
& (J2000) & (J2000) & (mag) & (mag) & (mag) & (mag) &&&(\AA)\\
\noalign{\smallskip}
\tableline
\noalign{\smallskip}

1 & 05 35 27.4 & -05 09 04 & 17.882 $\pm$ 0.036 & 14.852 $\pm$ 0.009 & 14.107 $\pm$ 0.008 & 13.594 $\pm$ 0.011 & M6.5\tablenotemark{d} &0.2&$-$19.2\\
2 & 05 35 16.8 & -05 07 27 & 17.819 $\pm$ 0.018 & 14.953 $\pm$ 0.009 & 14.413 $\pm$ 0.009 & 13.947 $\pm$ 0.011 & M6.5\tablenotemark{d} &0.0&$-$16.9\\
3 & 05 35 13.0 & -05 02 09 & 17.393 $\pm$ 0.005 & 14.465 $\pm$ 0.007 & 13.895 $\pm$ 0.007 & 13.477 $\pm$ 0.007 & M6.5\tablenotemark{d} &0.0&$-$25.8\\
29 & 05 35 12.9 & -05 15 49 & 18.447 $\pm$ 0.051 & 15.082 $\pm$ 0.015 & 14.309 $\pm$ 0.013 & 13.806 $\pm$ 0.015 & M8-M8.25\tablenotemark{e}, M8.5\tablenotemark{g} &0.2&neb\tablenotemark{j}\\
30 & 05 35 12.7 & -05 15 43 & 18.394 $\pm$ 0.069 & 14.825 $\pm$ 0.010 & 13.951 $\pm$ 0.010 & 13.328 $\pm$ 0.008 & M7 \tablenotemark{e}, M7.25\tablenotemark{g} &0.4&neb\tablenotemark{j}\\
31 & 05 35 18.7 & -05 15 18 & 18.003 $\pm$ 0.062 & 14.488 $\pm$ 0.009 & 13.658 $\pm$ 0.007 & 13.077 $\pm$ 0.009 & M7.75\tablenotemark{e}, M7.5\tablenotemark{h} &0.3&neb\tablenotemark{j}\\
21 & 05 35 38.2 & -05 03 34 & 18.561 $\pm$ 0.018 & 16.112 $\pm$ 0.027 & 15.373 $\pm$ 0.031 & 14.840 $\pm$ 0.024 & M7\tablenotemark{f}, M7.25\tablenotemark{h} &0.2&$-$105.1\\
34 & 05 35 13.6 & -05 14 22 & 18.445 $\pm$ 0.077 & 15.793 $\pm$ 0.020 & 14.880 $\pm$ 0.017 & 14.335 $\pm$ 0.022 & M8\tablenotemark{f} &0.4&\\
77 & 05 35 40.3 & -05 12 32 & 19.290 $\pm$ 0.40? & 15.462 $\pm$ 0.026 & 14.687 $\pm$ 0.018 & 14.155 $\pm$ 0.019 & M7\tablenotemark{f} &0.2&\\
923\_1621 & 05 35 15.1 & -04 58 09 & & 17.867 $\pm$ 0.051 & 16.197 $\pm$ 0.028 & 15.200 $\pm$ 0.029 & M8\tablenotemark{g} &1.5&\\
1036\_108 & 05 35 10.1 & -05 14 56 & & 16.551 $\pm$ 0.032 & 15.397 $\pm$ 0.022 & 14.594 $\pm$ 0.020 & M9\tablenotemark{g} &0.8&\\
543\_681&  05 35 32.0 & -05 08 35 & & 16.848 $\pm$ 0.026 & 15.430 $\pm$ 0.017 & 14.485 $\pm$ 0.014 & M7.25\tablenotemark{g} &1.2&\\
766\_341 & 05 35 22.1 & -05 12 21 & & 17.035 $\pm$ 0.081 & 15.908 $\pm$ 0.040 & 14.873 $\pm$ 0.027 & M7.75\tablenotemark{g} &0.7&\\
441\_326 & 05 35 36.6 & -05 12 31 & & 16.547 $\pm$ 0.070 & 15.758 $\pm$ 0.058 & 14.906 $\pm$ 0.051 & M8.5\tablenotemark{g} &0.3&\\
988\_148 & 05 35 12.2 & -05 14 29 & & 17.851 $\pm$ 0.069 & 16.503 $\pm$ 0.037 & 15.304 $\pm$ 0.030 & M8\tablenotemark{g} &1.1&\\
1004\_979 & 05 35 11.5 & -05 05 16 & & 16.321 $\pm$ 0.019 & 14.986 $\pm$ 0.011 & 14.330 $\pm$ 0.011 & M7\tablenotemark{g} &1.0&\\
727\_1506 & 05 35 23.9 & -04 59 25 & & 18.398 $\pm$ 0.070 & 16.800 $\pm$ 0.034 & 15.842 $\pm$ 0.031 & M7.75\tablenotemark{g} &1.4&\\
845\_236 & 05 35 23.3 & -05 13 48 & & 16.714 $\pm$ 0.060 & 15.789 $\pm$ 0.037 & 14.607 $\pm$ 0.029 & M7\tablenotemark{g} &0.5&\\
735\_36 & 05 35 28.2 & -05 16 01 &  & 15.790 $\pm$ 0.049 & 14.673 $\pm$ 0.020 & 13.978 $\pm$ 0.021 & M8-M8.25\tablenotemark{g} &0.7&\\

\noalign{\smallskip}
\tableline
\noalign{\smallskip}

\multicolumn{10}{l}{{$^a$ Coordinates from 2MASS final release.}}\\[1ex]
\multicolumn{10}{l}{{$^b$ Magnitudes from 4-Shooter photometry.}}\\[1ex]
\multicolumn{10}{l}{{$^c$ Magnitudes from SQIID photometry.}}\\[1ex]
\multicolumn{10}{l}{{$^d$ Keck LRIS spectra.}}\\[1ex]
\multicolumn{10}{l}{{$^e$ Blue Channel MMT spectra.}}\\[1ex]
\multicolumn{10}{l}{{$^f$ Red Channel MMT spectra.}}\\[1ex]
\multicolumn{10}{l}{{$^g$ SpeX IRTF spectra.}}\\[1ex]
\multicolumn{10}{l}{{$^h$ CorMASS spectra.}}\\[1ex]
\multicolumn{10}{l}{{$^i$ A$_H$ values calculated from $J-H$ colors and not from the A$_V$ values found as the best fit from spectral classification.}}\\[1ex]
\multicolumn{10}{l}{{$^j$ Too much emission from surrounding \ion{H}{II} region to measure H$\alpha$.}}\\[1ex]
\multicolumn{10}{l}{{$^k$ Table from \citet{peterson2007}.}}\\
\label{table:bdspectra}
\end{tabular}
}
\end{table}
\end{landscape}

\section{The NGC~1977 Region}

\begin{figure}[htb]
\plotfiddle{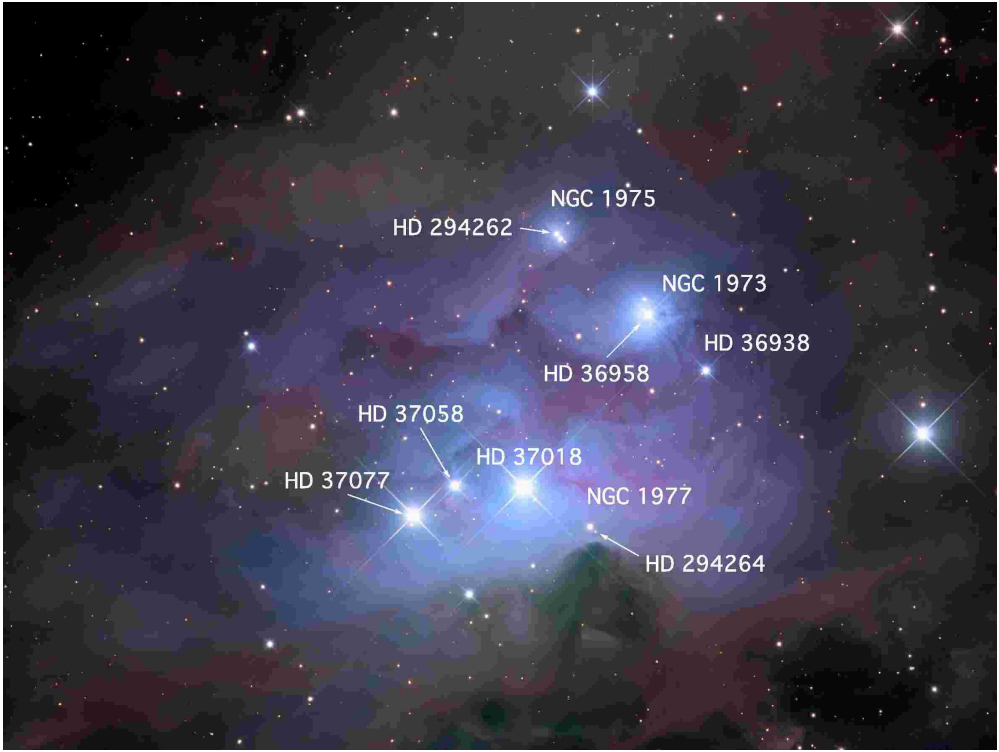}{9.5cm}{0}{22}{22}{-188}{0}
\caption[NGC~1977color]{The NGC~1977 HII region is primarily excited
by the B1~V star HD~37018. NGC~1973 and NGC~1975 are smaller nebulae,
but often the designation NGC~1977 is used for the whole
region. The field is approximately 30 $\times$ 45 arcmin, with north
up and east left.  Image courtesy Martin Pugh.}
\label{fig:ngc1977color}
\end{figure}

\begin{figure}[p]
\centering
\includegraphics[width=\textwidth,draft=False]{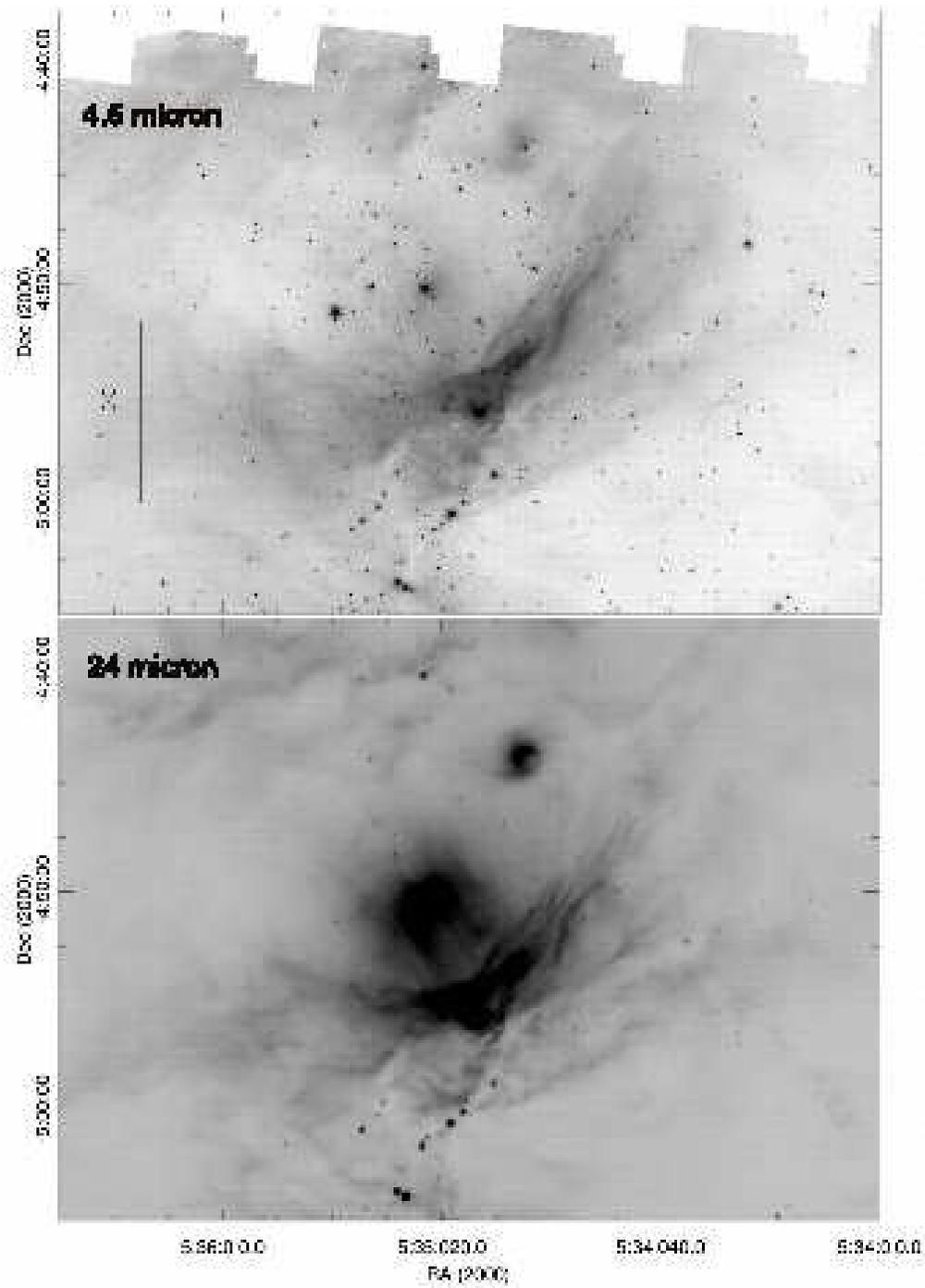}
\caption[NGC~1977]{NGC~1977 as seen at the {\it Spitzer} 4.5 and 24 $\mu$m
wavelengths \citep[][in prep]{megeath2007a}.  The bright star in the center
is the B1V star HD 37018.  Bright nebulosity is seen surrounding the star,
particularly at 24~$\mu$m, as well as south of this star at the edge of the
molecular filament.  Within this filament, protostars in the OMC-3 region can
be found (see Figure \ref{fig:ngc1977}).}
\label{fig:ngc1977bcd}
\end{figure}

The NGC~1977 nebula north of the Orion Nebula was discovered by
William Herschel in 1786, while the two smaller regions NGC~1973 and
1975 were noted by Heinrich Louis d'Arrest in 1862 and 1864,
respectively, see Figure~\ref{fig:ngc1977color}.  The designation
NGC~1977 is often used for the whole complex.  NGC~1977 comprises the
southern portion of the \ion{H}{II} region S279, at the northern end
of the Orion molecular cloud, defining an interface between the
\ion{H}{II} region and dense molecular cloud \citep*{ket76}.  The
ionizing source for the \ion{H}{II} region is the B1~V star HD~37018
(42~Ori) \citep{sharpless59,johnson77}.  In Figures
\ref{fig:ngc1977color} and \ref{fig:ngc1977bcd}, HD 37018 appears as the
brightest star near the middle of the image.  There are two B3 V stars that
flank the east (HD~37058) and south-west (HD~294264) sides of HD~37018; the
other star to the east of HD~37018, which appears almost as bright, is
HD~37077 (45~Ori), a F0~III star.  The star of approximately equal brightness
to the north-west is HD~36958 (KX~Ori), yet another B3~V star, which
illuminates NGC~1973.  Further to the northeast is HD~294262, an A0 double
star (the companion is LZ~Ori, also an A0 star), which illuminates NGC~1975.

Just south of the B3~V star HD~37058 is HH~45, the most prominent HH object in
NGC~1977.  It was first discovered by \citet{s77}, and is discussed in detail
in \citet{r89}.  The object has a bow-shock shape: the eastern side of the
bow-shock has a well defined boundary while the western side is more
extended.  The bow-shock has several bright knots, and an additional knot is
visible to the south of the bow-shock; these are shown in Figure 9b of
\citet{r89}.  The source for HH~45 is not known.  To the south-west of HH ~45,
\citet*{smz02} detected two knots of 2.12~$\mu$m H$_2$ emission
on either side of a near-infrared source; they refer to this source as SMZ~1.

\begin{figure}[ht!]
\centering
\includegraphics[width=\textwidth,draft=False]{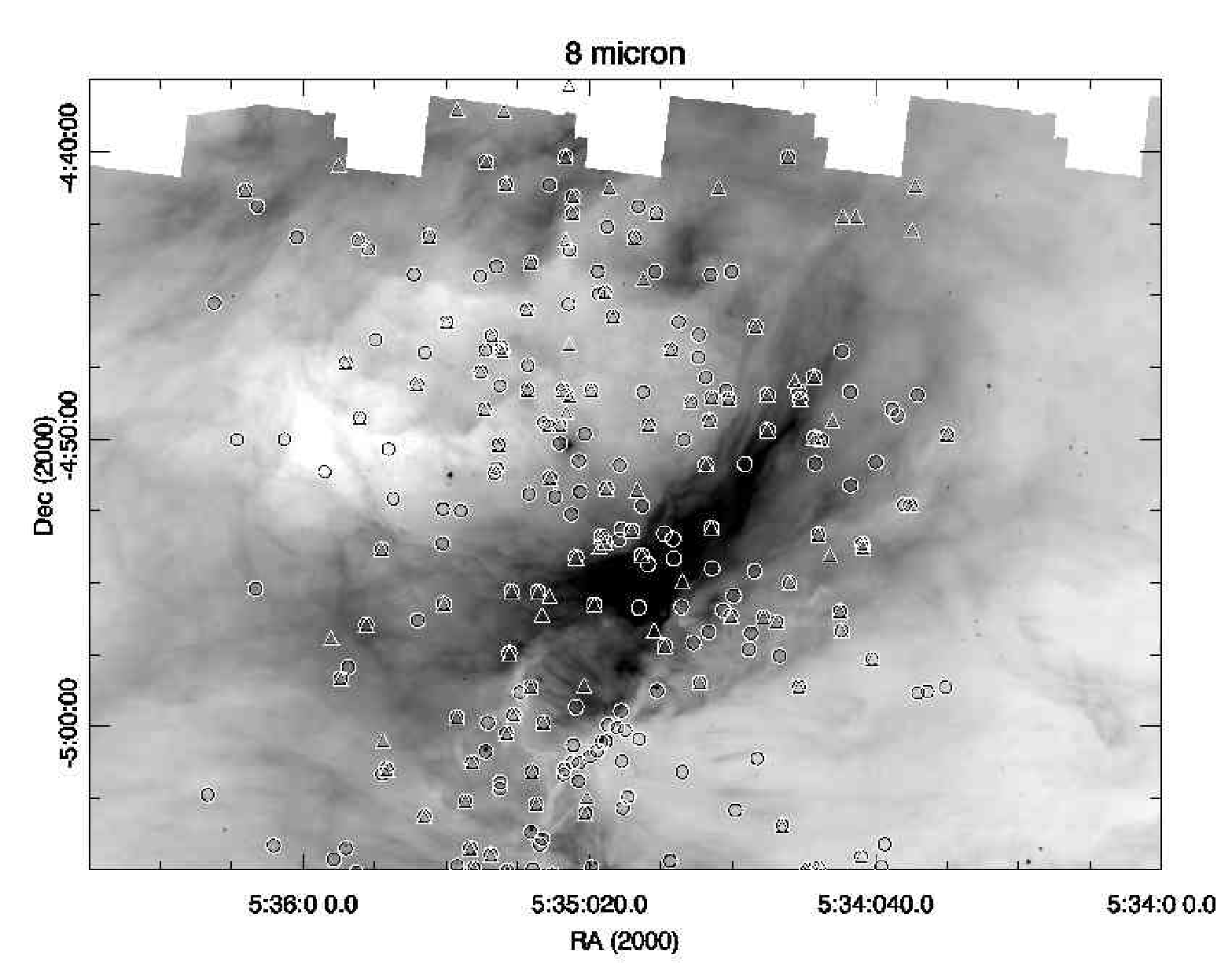}
\caption[NGC~1977 at 8 $\mu$m]{NGC~1977 as seen with {\it Spitzer} at 8 $\mu$m
\citep[][in prep]{megeath2007a}.  The bright star in the center is the B1V
star HD 37018.  The positions of young stars and protostars are
plotted: the triangles are variables from \citet*{chs01}, and the circles are
infrared excess sources from \citet[][in prep]{megeath2007a}.  A total of 170
young stellar objects are found with either infrared excess or variability.}
\label{fig:ngc1977}
\end{figure}

The {\it Spitzer} observations show that the \ion{H}{II} region fills
a loop of 8~$\mu$m emission (Figure~\ref{fig:ngc1977}). This suggests
that NGC~1977 is part of a larger cavity excavated in the molecular
gas by the B stars and that the \ion{H}{II} region is gas ionized off
the cavity by the B stars.  The 8~$\mu$m emission is brightest toward
NGC~1977, where the loop intersects the filament of dense molecular
gas and the OMC-2/3 region.  \citet{mhwe85} find that this region is
externally heated by HD~37018, and the radio continuum observations
presented in their paper rule out the possibility of additional
heating by any embedded stars earlier than B2.5-B3.  Detailed
observations of the interface between NGC~1977 and the molecular cloud
to the south by \citet{kmme85} show that the density of the molecular
cloud drops off gradually toward the interface.

The distribution of young stars in Figure~\ref{fig:ngc1977} shows a
concentration of stars within the loop.  These are presumably the young, low
mass stars associated with the B stars.  The {\it Spitzer} observations show
146 stars within NGC~1977 with infrared excesses, as selected by the
color-color diagrams in Figure~\ref{fig:tom_midircc}.  Of the 146 stars with
infrared excesses, six appear to be protostars (Class 0/I) based on
their 3.6$-$4.5 and 4.5$-$24 $\mu$m colors \citep[][in
prep]{megeath2007a}.  The remaining 140 infrared excess sources are
primarily young stars with disks (Class II), although some protostars
that were not detected at 24 $\mu$m may be among the 140 sources.  In
addition, \citet*{chs01} identified 97 variables in the same region;
24 of the variables do not have infrared excesses (Class III). In
total, there are 170 young stellar objects in NGC~1977 identified by
the presence of an infrared excess or variability (note that there is
no Chandra data toward NGC~1977 so X-rays were not used for pre-main
sequence star selection).  The positional limits used for this
analysis are $\alpha$ = 05$^h$ 34$^m$ 30$^s$ to 05$^h$ 36$^m$ 15$^s$,
$\delta$ = $-$04$^{\circ}$ 57$^{\prime}$ 00\arcsec to
$-$04$^{\circ}$ 37$^{\prime}$ 30\arcsec (J2000).  The
photometry for the young stellar objects in NGC~1977 with Spitzer
24~$\mu$m detections is presented in Table~\ref{table:pms_stars_1977}.

\section{Summary}

The northern end of the Orion A molecular cloud is the most active
region of low and high mass star formation within 500~pc of our Sun and
is one of the most intensively studied star forming regions in the sky.
However, most of these observations exclusively target the Orion Nebula and
the region immediately surrounding the nebula.  This chapter reviews the
rich concentration of pre-main sequence stars and protostars north of the
Orion Nebula in OMC-2/3 and NGC~1977.  These regions provide examples of
star formation in environments much different than that in the Orion Nebula; in
particular, in environments without the extreme-UV radiation found in the
Orion Nebula, and showing lower stellar densities than the densities
found in the center of the Orion Nebula.  Future studies of the ONC should
encompass OMC-2/3 and NGC~1977 in order to fully characterize the wide variety
of environments within this single contiguous star forming region and how
these environments affect star and planet formation.

\acknowledgments  We would like to thank Chris Davis for Figure
\ref{fig:outflows_chris} and for input into Section
\ref{sec:outflows}, Nick Siegler for providing the {\it Spitzer} 24
$\mu$m photometry, and Martin Pugh for the color image of NGC~1977.
The initial work on this chapter was done while co-author Peterson was at the
University of Virginia and while co-author Megeath was at the
Harvard-Smithsonian Center for Astrophysics.  Support for Megeath's work at
that time was provided by NASA through contract 1256970 issued by Jet
Propulsion Laboratory (JPL)/Caltech.  This work includes observations made
with the Spitzer Space Telescope, which is operated by JPL/Caltech under NASA
contract 1407.  This publication makes use of data products from the
Two-Micron All Sky Survey, which is a joint project of the University of
Massachusetts and the Infrared Processing and Analysis Center (IPAC)/Caltech,
funded by NASA and the NSF.  This research has made use of the NASA/IPAC
Infrared Science Archive, which is operated by JPL and Caltech under contract
with NASA, of the SIMBAD database, operated at CDS, Strasbourg, France, and of
NASA's Astrophysics Data System Bibliographic Services.

\setlength{\tabcolsep}{1.4\deftabcolsep}
\begin{landscape}
\begin{center}
{ \scriptsize
\begin{longtable}{ccccccccccc}

\caption{Young Stellar Object Candidates in NGC~1977 detected at 24 $\mu$m}\\
\noalign{\smallskip}
\hline
\noalign{\smallskip}
  RA &	Dec &	K$_S$\tablenotemark{a} & 3.6~$\mu$m & 4.5~$\mu$m & 5.8~$\mu$m & 8~$\mu$m & 24~$\mu$m & Var. & IR-ex\tablenotemark{b} & Proto.\\
  J(2000) & J(2000)  &	mag (unc) & mag (unc) & mag (unc) & mag (unc) & mag (unc) & mag (unc)  & & &\\
\noalign{\smallskip}
\hline
\noalign{\smallskip}
\endfirsthead

\caption{Young Stellar Object Candidates in NGC~1977 detected at 24~$\mu$m (continued)}\\
\noalign{\smallskip}
\hline
\noalign{\smallskip}
  RA &	Dec &	K$_S$\tablenotemark{a} & 3.6~$\mu$m & 4.5~$\mu$m & 5.8~$\mu$m & 8~$\mu$m & 24~$\mu$m & Var. & IR-ex\tablenotemark{b} & Proto.\\
  J(2000) & J(2000)  &	mag (unc) & mag (unc) & mag (unc) & mag (unc) & mag (unc) & mag (unc)  & & &\\
\noalign{\smallskip}
\hline
\noalign{\smallskip}
\endhead

\noalign{\smallskip}
\hline
\noalign{\smallskip}
\multicolumn{10}{l}{{$^a$ Magnitudes from 2MASS photometry.}}\\[1ex]
\multicolumn{10}{l}{{$^b$ From \citet*{chs01}; Those that are variable have Stetson indices greater than 0.55.}}\\
\endfoot

\noalign{\smallskip}
\hline
\noalign{\smallskip}
\multicolumn{10}{l}{{$^a$ Magnitudes from 2MASS photometry.}}\\[1ex]
\multicolumn{10}{l}{{$^b$ From \citet*{chs01}; Those that are variable have Stetson indices greater than 0.55.}}\\
\endlastfoot

 5 34 57.4 & -4 56 45 &  11.63 (0.02) & 10.42 (0.01) &  9.81 (0.01) &  9.23 (0.01) &  8.16 (0.01) &  5.73 (0.02) & 0 &  1 &  0 \\
 5 35 03.3 & -4 56 42 &  11.31 (0.02) & 10.92 (0.01) & 10.57 (0.01) & 10.34 (0.03) &  9.47 (0.04) &  5.83 (0.02) & 0 &  1 &  0 \\
 5 34 44.8 & -4 56 40 &  12.23 (0.02) & 11.65 (0.01) & 11.11 (0.01) & 10.89 (0.02) & 10.38 (0.05) &  7.53 (0.08) & 1 &  1 &  0 \\
 5 35 51.2 & -4 56 27 &  11.54 (0.02) & 11.14 (0.01) & 10.50 (0.01) &  9.81 (0.02) &  8.66 (0.03) &  4.81 (0.08) & 1 &  1 &  0 \\
 5 34 53.9 & -4 56 22 &  10.91 (0.02) & 10.22 (0.01) & 10.33 (0.01) &  9.81 (0.01) &  9.36 (0.02) &  4.87 (0.01) & 1 &  1 &  0 \\
 5 34 55.7 & -4 56 12 &  10.22 (0.02) &  9.22 (0.01) &  8.83 (0.01) &  8.50 (0.01) &  7.79 (0.01) &  4.71 (0.04) & 1 &  1 &  0 \\
 5 35 00.3 & -4 56 08 &  12.57 (0.02) & 11.81 (0.01) & 11.48 (0.01) & 11.19 (0.04) & 10.51 (0.07) &  6.78 (0.01) & 1 &  1 &  0 \\
 5 34 45.0 & -4 55 59 &  11.35 (0.03) & 10.18 (0.01) &  9.78 (0.01) &  9.53 (0.01) &  8.96 (0.03) &  5.61 (0.05) & 1 &  1 &  0 \\
 5 35 19.3 & -4 55 44 &  10.08 (0.02) &  8.28 (0.01) &  7.27 (0.01) &  6.70 (0.01) &  5.94 (0.01) &  2.66 (0.06) & 1 &  1 &  0 \\
 5 35 40.4 & -4 55 44 &  12.20 (0.02) & 11.70 (0.01) & 11.55 (0.01) & 11.42 (0.03) & 10.66 (0.04) &  6.35 (0.08) & 1 &  1 &  0 \\
 5 35 30.9 & -4 55 17 &  10.67 (0.02) & 10.13 (0.01) &  9.75 (0.01) &  9.41 (0.02) &  8.36 (0.03) &  4.72 (0.17) & 1 &  1 &  0 \\
 5 34 52.0 & -4 55 00 &  11.69 (0.02) & 10.89 (0.01) & 10.47 (0.01) & 10.18 (0.02) &  9.14 (0.02) &  5.49 (0.06) & 1 &  1 &  0 \\
 5 35 08.2 & -4 54 09 &  10.15 (0.02) &  8.16 (0.01) &  7.36 (0.01) &  6.43 (0.01) &  5.06 (0.01) &  2.48 (0.03) & 0 &  0 &  1 \\
 5 34 41.8 & -4 53 46 &  11.03 (0.02) & 10.54 (0.01) & 10.24 (0.01) &  9.95 (0.02) &  8.98 (0.02) &  5.40 (0.02) & 1 &  1 &  0 \\
 5 35 40.5 & -4 53 37 &    & 15.13 (0.07) & 14.98 (0.09) & 12.16 (0.09) & 10.32 (0.09) &  6.42 (0.10) & 0 &  1 &  0 \\
 5 35 15.9 & -4 53 29 &  13.52 (0.03) & 12.52 (0.01) & 12.10 (0.02) & 11.56 (0.08) &   &  6.00 (0.03) & 0 &  1 &  0 \\
 5 35 18.4 & -4 53 23 &  10.11 (0.02) &  9.01 (0.01) &  8.47 (0.01) &  8.09 (0.01) &  7.28 (0.01) &  4.19 (0.01) & 1 &  1 &  0 \\
 5 35 14.1 & -4 53 11 &  12.34 (0.02) & 10.65 (0.01) & 10.10 (0.01) &  9.56 (0.01) &  8.82 (0.01) &  4.82 (0.04) & 1 &  1 &  0 \\
 5 35 22.5 & -4 52 36 &  10.68 (0.02) &  9.72 (0.01) &  9.09 (0.01) &  8.44 (0.01) &  7.24 (0.01) &  4.15 (0.19) & 0 &  1 &  0 \\
 5 35 40.5 & -4 52 26 &  11.62 (0.02) & 11.25 (0.01) & 10.98 (0.01) & 10.79 (0.02) & 10.14 (0.02) &  7.38 (0.02) & 0 &  1 &  0 \\
 5 34 35.2 & -4 52 17 &  11.80 (0.02) & 10.66 (0.01) &  9.85 (0.01) &  9.04 (0.01) &  7.92 (0.01) &  3.66 (0.01) & 1 &  0 &  1 \\
 5 35 47.4 & -4 52 04 &  15.04 (0.10) & 14.35 (0.03) & 13.75 (0.03) & 13.24 (0.09) & 12.34 (0.08) &  7.85 (0.04) & 0 &  1 &  0 \\
 5 35 13.3 & -4 51 44 &   7.58 (0.03) &  7.47 (0.01) &  7.42 (0.01) &  7.40 (0.01) &  7.36 (0.01) &  4.42 (0.05) & 1 &  0 &  0 \\
 5 34 43.5 & -4 51 36 &  11.77 (0.02) & 11.64 (0.01) & 11.59 (0.01) & 11.34 (0.05) & 10.77 (0.12) &  6.37 (0.02) & 0 &  1 &  0 \\
 5 35 25.5 & -4 51 20 &  10.69 (0.02) &  9.98 (0.01) &  9.45 (0.01) &  9.25 (0.01) &  8.20 (0.01) &  4.75 (0.05) & 1 &  1 &  0 \\
 5 35 03.7 & -4 50 52 &   9.60 (0.02) &  8.92 (0.01) &  8.41 (0.01) &  8.12 (0.01) &  7.25 (0.02) &  3.75 (0.02) & 1 &  1 &  0 \\
 5 34 48.4 & -4 50 51 &  12.38 (0.02) & 11.81 (0.01) & 11.36 (0.01) & 11.02 (0.03) & 10.14 (0.05) &  6.53 (0.05) & 0 &  1 &  0 \\
 5 35 21.5 & -4 50 45 &   8.94 (0.02) &  8.30 (0.01) &  8.08 (0.01) &  7.83 (0.01) &  7.11 (0.01) &  2.78 (0.02) & 0 &  1 &  0 \\
 5 35 48.1 & -4 50 20 &  12.30 (0.02) & 11.87 (0.01) & 11.69 (0.01) & 11.50 (0.03) & 10.93 (0.03) &  7.25 (0.04) & 0 &  1 &  0 \\
 5 35 32.7 & -4 50 11 &   9.62 (0.03) &  8.73 (0.01) &  8.30 (0.01) &  7.99 (0.01) &  7.33 (0.01) &  4.69 (0.06) & 1 &  1 &  0 \\
 5 36 09.3 & -4 50 00 &  11.80 (0.02) & 11.23 (0.01) & 10.94 (0.01) & 10.72 (0.02) & 10.00 (0.01) &  7.24 (0.04) & 0 &  1 &  0 \\
 5 35 06.8 & -4 50 01 &  15.40 (0.16) & 14.51 (0.03) & 13.65 (0.03) & 12.63 (0.06) & 10.94 (0.05) &  6.73 (0.01) & 0 &  0 &  1 \\
 5 34 47.6 & -4 50 01 &  10.95 (0.02) & 10.58 (0.01) & 10.47 (0.01) & 10.38 (0.03) &  9.37 (0.03) &  4.23 (0.15) & 1 &  1 &  0 \\
 5 34 48.5 & -4 49 56 &  11.16 (0.02) &  9.91 (0.01) &  9.30 (0.01) &  9.36 (0.02) &  8.58 (0.04) &  5.30 (0.03) & 1 &  1 &  0 \\
 5 34 30.1 & -4 49 50 &  10.96 (0.02) & 10.51 (0.01) & 10.33 (0.01) & 10.31 (0.02) & 10.06 (0.02) &  6.34 (0.01) & 1 &  1 &  0 \\
 5 35 11.8 & -4 49 30 &  12.11 (0.02) & 11.02 (0.01) & 10.50 (0.01) & 10.09 (0.01) &  9.28 (0.01) &  5.44 (0.06) & 1 &  1 &  0 \\
 5 35 03.2 & -4 49 20 &   8.51 (0.02) &  7.68 (0.01) &  7.25 (0.01) &  7.08 (0.01) &  6.59 (0.01) &  3.86 (0.02) & 1 &  1 &  0 \\
 5 35 52.1 & -4 49 15 &  12.61 (0.02) & 11.97 (0.01) & 11.48 (0.01) & 11.31 (0.03) & 10.61 (0.02) &  8.33 (0.03) & 1 &  1 &  0 \\
 5 34 37.0 & -4 49 10 &  12.89 (0.02) & 11.94 (0.01) & 11.63 (0.01) & 11.44 (0.03) & 10.92 (0.06) &  7.70 (0.06) & 0 &  1 &  0 \\
 5 35 34.6 & -4 48 57 &  12.58 (0.03) & 10.73 (0.01) & 10.12 (0.01) &  9.35 (0.01) &  8.40 (0.01) &  5.54 (0.02) & 1 &  1 &  0 \\
 5 34 37.7 & -4 48 57 &  11.91 (0.02) & 11.75 (0.01) & 11.70 (0.01) & 11.65 (0.03) & 11.34 (0.06) &  6.56 (0.07) & 0 &  1 &  0 \\
 5 35 05.8 & -4 48 42 &  10.57 (0.02) &  9.88 (0.01) &  9.52 (0.01) &  9.37 (0.01) &  8.62 (0.01) &  5.88 (0.03) & 1 &  1 &  0 \\
 5 34 50.6 & -4 48 37 &  12.64 (0.02) & 11.49 (0.01) & 10.95 (0.01) & 10.58 (0.03) &  9.77 (0.06) &  6.60 (0.29) & 1 &  1 &  0 \\
 5 35 00.5 & -4 48 35 &  12.37 (0.02) & 11.23 (0.01) & 10.64 (0.01) & 10.19 (0.02) &  9.35 (0.03) &  6.39 (0.02) & 1 &  1 &  0 \\
 5 35 03.0 & -4 48 32 &  12.19 (0.03) & 11.78 (0.01) & 11.54 (0.01) & 11.18 (0.03) & 10.02 (0.03) &  6.88 (0.01) & 1 &  1 &  0 \\
 5 34 34.2 & -4 48 27 &  11.71 (0.02) & 11.40 (0.01) & 11.23 (0.01) & 11.03 (0.03) & 10.21 (0.05) &  6.54 (0.01) & 0 &  1 &  0 \\
 5 35 23.2 & -4 48 27 &  14.50 (0.09) & 13.18 (0.02) & 12.77 (0.02) & 12.26 (0.05) & 10.89 (0.04) &  5.47 (0.07) & 0 &  1 &  0 \\
 5 34 55.1 & -4 48 27 &  10.17 (0.02) & 10.08 (0.01) & 10.07 (0.01) & 10.04 (0.02) &  9.92 (0.09) &  4.47 (0.03) & 1 &  1 &  0 \\
 5 35 00.9 & -4 48 18 &  11.26 (0.02) & 10.79 (0.01) & 10.50 (0.01) & 10.19 (0.02) &  9.37 (0.02) &  6.98 (0.01) & 0 &  1 &  0 \\
 5 35 23.7 & -4 48 16 &  12.02 (0.02) & 11.53 (0.01) & 11.10 (0.01) & 10.60 (0.01) &  9.33 (0.01) &  6.07 (0.01) & 1 &  1 &  0 \\
 5 35 28.7 & -4 48 16 &  12.88 (0.02) &  9.40 (0.01) &  8.65 (0.01) &  7.93 (0.01) &  7.13 (0.01) &  3.87 (0.02) & 1 &  1 &  0 \\
 5 35 44.1 & -4 48 04 &  10.72 (0.02) & 10.01 (0.01) &  9.70 (0.01) &  9.58 (0.01) &  9.04 (0.01) &  5.75 (0.03) & 1 &  1 &  0 \\
 5 34 48.6 & -4 47 50 &  11.94 (0.02) & 11.47 (0.01) & 10.95 (0.01) & 10.86 (0.03) & 10.07 (0.03) &  6.51 (0.02) & 1 &  1 &  0 \\
 5 35 35.2 & -4 47 39 &  11.26 (0.02) & 10.80 (0.01) & 10.52 (0.01) & 10.17 (0.02) &  8.81 (0.01) &  4.49 (0.01) & 1 &  1 &  0 \\
 5 35 28.6 & -4 47 26 &  11.20 (0.02) & 10.63 (0.01) & 10.25 (0.01) &  9.91 (0.01) &  9.03 (0.02) &  5.95 (0.07) & 0 &  1 &  0 \\
 5 35 54.1 & -4 47 19 &  11.04 (0.02) & 10.05 (0.01) &  9.60 (0.01) &  9.23 (0.01) &  8.63 (0.01) &  5.65 (0.02) & 1 &  1 &  0 \\
 5 35 43.0 & -4 47 00 &  12.18 (0.02) & 11.87 (0.01) & 11.66 (0.01) & 11.39 (0.03) & 10.74 (0.04) &  8.32 (0.11) & 0 &  1 &  0 \\
 5 35 32.2 & -4 46 57 &  10.17 (0.02) &  9.41 (0.01) &  9.00 (0.01) &  8.62 (0.01) &  7.66 (0.01) &  4.72 (0.01) & 1 &  1 &  0 \\
 5 35 34.5 & -4 46 54 &  11.01 (0.02) & 10.67 (0.01) & 10.22 (0.01) &  9.66 (0.01) &  8.64 (0.01) &  5.71 (0.04) & 1 &  1 &  0 \\
 5 35 08.7 & -4 46 52 &   9.91 (0.02) &  8.64 (0.01) &  8.28 (0.01) &  7.73 (0.01) &  7.04 (0.01) &  4.23 (0.01) & 1 &  1 &  0 \\
 5 35 32.3 & -4 46 48 &  13.57 (0.06) & 11.66 (0.01) & 10.84 (0.01) & 10.15 (0.02) &  9.31 (0.02) &  5.30 (0.08) & 1 &  0 &  1 \\
 5 35 22.8 & -4 46 41 &  11.80 (0.02) & 11.23 (0.01) & 11.02 (0.01) & 10.68 (0.02) & 10.11 (0.02) &  6.93 (0.30) & 1 &  0 &  0 \\
 5 35 49.9 & -4 46 32 &  11.46 (0.02) & 11.06 (0.01) & 10.75 (0.01) & 10.41 (0.02) &  9.68 (0.03) &  6.63 (0.01) & 0 &  1 &  0 \\
 5 35 33.7 & -4 46 23 &  10.84 (0.02) & 10.47 (0.01) & 10.35 (0.01) & 10.23 (0.02) &  9.67 (0.02) &  5.33 (0.07) & 1 &  1 &  0 \\
 5 35 04.7 & -4 46 21 &  13.38 (0.03) & 12.96 (0.02) & 12.51 (0.02) & 12.10 (0.05) & 11.51 (0.08) &  7.82 (0.07) & 0 &  1 &  0 \\
 5 34 56.8 & -4 46 04 &  10.96 (0.02) &  9.94 (0.01) &  9.64 (0.01) &  9.25 (0.01) &  9.00 (0.02) &  6.54 (0.02) & 1 &  1 &  0 \\
 5 35 40.0 & -4 45 54 &  10.42 (0.02) & 10.03 (0.01) &  9.79 (0.01) &  9.48 (0.01) &  8.62 (0.01) &  6.02 (0.01) & 1 &  1 &  0 \\
 5 35 16.7 & -4 45 44 &   9.72 (0.02) &  8.29 (0.01) &  8.01 (0.01) &  7.60 (0.01) &  6.80 (0.01) &  4.42 (0.03) & 1 &  1 &  0 \\
 5 35 28.8 & -4 45 29 &  12.74 (0.03) & 11.22 (0.01) & 10.62 (0.01) & 10.08 (0.02) &  9.08 (0.01) &  6.12 (0.01) & 1 &  1 &  0 \\
 5 35 22.9 & -4 45 18 &  12.77 (0.03) & 12.31 (0.01) & 12.36 (0.01) & 11.96 (0.04) & 10.95 (0.04) &  7.08 (0.03) & 0 &  1 &  0 \\
 5 36 12.4 & -4 45 15 &    & 15.46 (0.08) & 14.68 (0.07) & 12.63 (0.09) & 10.98 (0.11) &  7.43 (0.06) & 0 &  0 &  1 \\
 5 35 17.9 & -4 44 52 &  10.16 (0.02) &  9.59 (0.01) &  9.13 (0.01) &  9.16 (0.01) &  8.16 (0.01) &  4.84 (0.07) & 1 &  1 &  0 \\
 5 35 44.5 & -4 44 15 &  11.45 (0.02) & 11.34 (0.01) & 11.28 (0.01) & 11.20 (0.04) & 10.57 (0.08) &  6.43 (0.07) & 0 &  1 &  0 \\
 5 35 18.9 & -4 44 10 &  11.91 (0.02) & 11.43 (0.01) & 11.25 (0.01) & 10.96 (0.04) & 10.28 (0.07) &  7.53 (0.08) & 0 &  1 &  0 \\
 5 35 33.0 & -4 43 58 &  10.41 (0.02) & 10.01 (0.01) &  9.79 (0.01) &  9.56 (0.01) &  8.56 (0.02) &  4.10 (0.05) & 0 &  1 &  0 \\
 5 35 22.8 & -4 43 25 &  12.11 (0.05) & 11.62 (0.01) & 11.29 (0.01) & 10.85 (0.02) &  9.51 (0.02) &  5.71 (0.05) & 1 &  1 &  0 \\
 5 35 52.3 & -4 43 04 &  10.85 (0.02) & 10.56 (0.01) & 10.56 (0.01) & 10.44 (0.03) & 10.46 (0.05) &  6.00 (0.10) & 1 &  1 &  0 \\
 5 35 13.7 & -4 42 58 &  10.64 (0.02) & 10.24 (0.01) & 10.01 (0.01) &  9.88 (0.02) &  9.22 (0.02) &  6.75 (0.11) & 1 &  1 &  0 \\
 5 35 42.4 & -4 42 56 &  14.21 (0.04) & 12.84 (0.03) & 12.31 (0.02) & 11.32 (0.05) &  9.95 (0.04) &  6.42 (0.02) & 1 &  1 &  0 \\
 5 35 17.5 & -4 42 37 &    & 12.93 (0.02) & 12.27 (0.02) & 11.71 (0.05) & 10.93 (0.04) &  8.40 (0.02) & 0 &  1 &  0 \\
 5 35 22.3 & -4 42 07 &  10.14 (0.02) &  9.47 (0.01) &  9.27 (0.01) &  9.01 (0.01) &  8.12 (0.03) &  4.35 (0.07) & 1 &  1 &  0 \\
 5 36 06.4 & -4 41 53 &  10.71 (0.02) & 10.30 (0.01) & 10.06 (0.01) &  9.64 (0.02) &  8.79 (0.04) &  5.41 (0.02) & 0 &  1 &  0 \\
 5 35 13.2 & -4 41 54 &  11.80 (0.02) & 11.31 (0.01) & 10.87 (0.01) & 10.52 (0.03) &  9.76 (0.03) &  6.91 (0.07) & 0 &  1 &  0 \\
 5 36 08.1 & -4 41 20 &  10.71 (0.02) & 10.34 (0.01) &  9.95 (0.01) &  9.46 (0.02) &  8.63 (0.03) &  5.25 (0.06) & 1 &  1 &  0 \\
 5 35 25.6 & -4 41 08 &  12.12 (0.02) & 11.55 (0.01) & 11.26 (0.01) & 10.89 (0.04) & 10.37 (0.08) &  7.42 (0.03) & 0 &  1 &  0 \\
 5 35 31.7 & -4 41 07 &  12.18 (0.02) & 11.85 (0.01) & 11.73 (0.02) & 11.63 (0.05) & 11.27 (0.07) &  6.10 (0.03) & 1 &  1 &  0 \\
 5 35 34.5 & -4 40 20 &  10.60 (0.02) & 10.08 (0.01) & 10.20 (0.01) &  9.75 (0.03) &  9.50 (0.10) &  4.31 (0.06) & 1 &  1 &  0 \\
 5 35 23.3 & -4 40 10 &   9.53 (0.02) &  7.33 (0.01) &  6.25 (0.01) &  5.86 (0.01) &  4.66 (0.01) &  1.46 (0.08) & 1 &  0 &  1 \\
 5 34 52.2 & -4 40 11 &   9.80 (0.02) &  9.07 (0.01) &  8.64 (0.01) &  8.58 (0.01) &  7.93 (0.01) &  5.15 (0.06) & 1 &  1 &  0 \\
 5 35 38.5 & -4 38 32 &  12.34 (0.02) &   &   &   &   &  7.63 (0.06) & 1 &  0 &  0 \\
 5 35 22.8 & -4 37 39 &    &   &   &   &   &  5.94 (0.27) & 1 &  0 &  0\\


\label{table:pms_stars_1977}
\end{longtable}
}
\end{center}
\end{landscape}
\setlength{\tabcolsep}{\deftabcolsep}

\end{document}